\documentclass[aps,preprint,nofootinbib]{revtex4}

\usepackage{graphicx}
\usepackage{amssymb}
\usepackage{amsmath}
\usepackage{epstopdf}
\usepackage{ulem}

\graphicspath{{/Users/sofia/lensing/lensing_ergasies/ergasia_wmap/paper_REV3/}}

\usepackage[greek,english]{babel}
\usepackage[iso-8859-7]{inputenc}

\usepackage{color}

\def\ba{\begin{eqnarray}}
\def\ea{\end{eqnarray}}
\def\be{\begin{equation}}
\def\ee{\end{equation}}

\def\gtorder{\mathrel{\raise.3ex\hbox{$>$}\mkern-14mu
             \lower0.6ex\hbox{$\sim$}}}
\def\ltorder{\mathrel{\raise.3ex\hbox{$<$}\mkern-14mu
             \lower0.6ex\hbox{$\sim$}}}

\def\dalemb#1#2{{\vbox{\hrule height.#2pt
  \hbox{\vrule width.#2pt height#1pt \kern#1pt \vrule width.#2pt}
    \hrule height.#2pt}}}

\begin{document}

\rightline{T$\Theta\Delta$}

\title{CMB lensing reconstruction from the WMAP 7--year data}
\author{C. Sofia Carvalho$^{1,2}$}\email{cscarvalho@oal.ul.pt}
\author{Ismael Tereno$^{1}$}
\author{Spyros Basilakos$^{2}$}
\address{$^1$ 
Centro de Astronomia e Astrof\'isica da Universidade de Lisboa, Tapada da Ajuda, 1349-018 Lisboa, Portugal}
\address{$^2$
Research Center for Astronomy and Applied Mathematics, Academy of Athens, Soranou Efessiou 4, GR 11 527 Athens, Greece}
\begin{abstract}
We attempt to make a direct measurement of the weak lensing signal from the WMAP 7--year data. We apply the real--space implementation of the optimal quadratic estimator on the maps produced by the W--band Differencing Assemblies. We obtain a weak lensing amplitude parameter of $A_{L}=0.99\pm1.67$ after correcting for several sources of bias.
The error budget includes a contribution from the bias removal procedure.
Despite the demonstrated insensitivity of the real--space estimator to uncorrelated noise, we conclude that this detection is not statistically significant. 
We expect that a full--sky, higher--sensitivity experiment such as Planck will allow us to make a more significant measurement.
\end{abstract}
\date{\today}
\maketitle

\section{Introduction}


Gravitational lensing of the cosmic microwave background (CMB) radiation is a source of information on the interaction between the CMB photons and the fluctuations in the matter density distributed along their trajectory. 
This interaction causes the deflection of the CMB photons by the gradient of the lensing potential $\psi,$ which results in the creation of correlations among initially uncorrelated modes.
Since the matter fluctuations are in the linear regime on large scales, the deflections are small and consequently the total deflection can be described as a sum of multiple deflections along the unperturbed trajectory (Born approximation). The correlations appear in the form of non--diagonal couplings between gradients of the temperature and gradients of the lensing potential, hence as a non--linear signal which becomes important at small scales.
By measuring these correlations, we can estimate the lensing deflection  
${\boldsymbol \alpha}=\nabla\psi$
and consequently the lensing convergence 
${\boldsymbol \kappa}=-\nabla^2\psi/2$ 
which is a direct measure of the projected matter density.


Detections of a lensing signal in the CMB are usually summarized in the lensing amplitude parameter $A_{L}$ which describes the amplitude of the detected lensing power spectrum relative to the theoretical prediction. Thus, $A_{L}=0$ indicates no lensing, while $A_{L}=1$ indicates the standard $\Lambda$CDM model and General Relativity lensing theory.    
Recently there were reported direct detections of the lensing signal from experiments designed to look for non-linear signals in the CMB, yielding $A_{L}=1.16\pm0.29$ from the Atacama Cosmology Telescope (ACT) data  \cite{das11},
and $A_{L}=0.90\pm 0.19$ from the South Pole Telescope (SPT) data  \cite{engelen12}. 
Although the Wilkinson Microwave Anisotropy Probe (WMAP) was not designed to look for non--linear signals, a detection from the WMAP full--sky data would provide a full--sky measurement of the lensing signal and complement the other detections. 

We anticipate some difficulties in this detection.
Since the experimental noise in the WMAP data overwhelms the lensing signal, a statistically significant measurement should not be possible and indeed the first detections were made indirectly by correlating the results with galaxy counts \cite{smith07, hirata08}. There is however a reported direct detection of the lensing signal from the WMAP data using a kurtosis estimator, finding $A_{L}=\{0.96\pm0.60,1.06\pm0.69,0.97\pm0.47\}$ from the data of the V, W and combined V+W bands respectively \cite{smidt11}, whereas a more recent less significant direct detection from WMAP data, using the optimal quadratic estimator \cite{hu01}, found $A_{L}=1.27\pm0.98$ \cite{feng11}.

Here we base our measurement on an estimator with demonstrated insensitivity to uncorrelated noise \cite{moodley} and good handling of excision of points \cite{tereno}, which makes it a reliable tool for the extraction of the lensing signal from the WMAP data. 
Similarly to the optimal estimator, this estimator combines two filters of the temperature map weighted by the inverse of the variance. Instead of acting in Fourier space and using all the available sky for the reconstruction of the lensing potential at any point, this estimator acts in real space and uses the neighbouring points only, which presupposes a truncation of the kernel that convolves the two filters. Although truncating the kernel renders the estimator slightly suboptimal, it has a strong physical motivation.
Since weak lensing is a small--scale effect coherent over scales of a few degrees \cite{seljak96}, information on the lensing potential at a given point is contained in the points within the degree scale, so we expect little gain in considering the information over the entire sky \cite{bucher}. 
Thus a direct detection of a lensing signal from the WMAP data is both a test of the robustness of the real--space estimator and a comparison of the performance with other estimators. 

The manuscript is organized as follows.
In Sec.~\ref{sec:wmap_begin} we introduce the weak lensing estimator. In Sec.~\ref{sec:wmap_data} we justify our choice of data and describe the simulations. We then proceed to measure the weak lensing signal. In Sec.~\ref{sec:wmap_ps} we use synthesised map so as to simulate a control experiment, test the bias removal and setup the calibration.
In Sec.~\ref{sec:wmap_maps} we use patches from cartesian projections of the full--sky maps. The results are summarized in Sec.~\ref{sec:wmap_end}.

\section{Estimator of the CMB weak lensing}
\label{sec:wmap_begin}

The deflection of the CMB photons by the 
gradient of the lensing potential 
amounts to remapping the unlensed temperature anisotropies $T$ to the lensed $\tilde T$ in the direction ${\boldsymbol \theta}$ by $\tilde T({\boldsymbol \theta})=T({\boldsymbol \theta}+\nabla\psi),$ where the lensing potential $\psi$ is the projection on the image plane of the gravitational potential of the large--scale structure. For small deflections, the lensed temperature power spectrum is
\ba
\big< \tilde T({\boldsymbol \ell}^{\prime})\tilde T(\boldsymbol{\ell-\ell}^{\prime})\big> 
=(2\pi)^2\delta({\boldsymbol \ell})C_{\ell^{\prime}}
+(2\pi)^2\left[ 
{\boldsymbol \ell}\cdot {\boldsymbol \ell}^{\prime} C_{\ell^{\prime}}
+
{\boldsymbol \ell}\cdot ({\boldsymbol \ell}-{\boldsymbol \ell}^{\prime}) C_{\vert {\boldsymbol \ell}-{\boldsymbol \ell}^{\prime}\vert}
\right]\psi({\boldsymbol \ell}) 
+O(\psi^2),
\label{eqn:cl_TT_lensed}
\ea
where $C_{\ell}$ is the unlensed temperature power spectrum. 
Both the unlensed temperature anisotropies and the large--scale structure are assumed to be Gaussian random fields. The remapping of one Gaussian field by another induces non--Gaussianities.
The lensing signal is encoded in the off--diagonal correlations as a coupling between the photons and the gradient of the lensing potential. 
An estimator that captures the information in the coupling consists of a quadratic combination of the temperature map optimized so as to yield the minimum variance \cite{hu01}.
This estimator will also pick up the diagonal correlations which exist in the absence of lensing and which we must remove both from the estimated map and from the computed power spectrum. The diagonal correlations have a much larger amplitude than the lensing signal and are computed from simulated unlensed temperature maps. 


We measure the isotropic component of the convergence tensor, defined as $\kappa =-(\partial_{x}^2+\partial_{y}^2)\psi/2.$ We use the real--space estimator introduced in Ref.~\cite{moodley} which is based on the convolution of the square of the lensed  temperature map $\tilde T$ with a local kernel $W$ as follows
\ba
\hat \kappa (\boldsymbol\theta)&=&
\int d^2\boldsymbol\theta_{+}\int d^2\boldsymbol\theta_{-}
\tilde T(\boldsymbol\theta_{+}+\boldsymbol\theta_{-})
\tilde T(\boldsymbol\theta_{+}-\boldsymbol\theta_{-})
W(\boldsymbol\theta,\boldsymbol\theta_{+},\boldsymbol\theta_{-}).
\ea
The real--space kernel $W(\boldsymbol\theta,\boldsymbol\theta_{+},\boldsymbol\theta_{-})$ is given by 
\ba
W(\boldsymbol\theta,\boldsymbol\theta_{+},\boldsymbol\theta_{-})
=\int {d^2{\boldsymbol \ell_{+}}\over {(2\pi)^2}}
\exp\left[i{\boldsymbol\ell_{+}}\cdot {\boldsymbol\theta} \right]
\int {d^2{\boldsymbol \ell_{-}}\over {(2\pi)^2}}
\exp\left[-i({\boldsymbol\ell_{+}}\cdot {\boldsymbol\theta_{+}}
+{\boldsymbol\ell_{-}}\cdot {\boldsymbol\theta_{-}})\right]
W({\boldsymbol\ell_{+}},{\boldsymbol\ell_{-}})
\ea
where the corresponding Fourier--space kernel $W({\boldsymbol\ell_{+}},{\boldsymbol\ell_{-}})$ is related to the minimum--variance (to leading order) weight function  $Q(\boldsymbol{\ell},\boldsymbol{\ell}^{\prime})$ by
\ba
W({\boldsymbol\ell_{+}},{\boldsymbol\ell_{-}})
=Q(\boldsymbol{\ell},\boldsymbol{\ell}^{\prime})
\ea
upon the coordinate transformation ${\boldsymbol \ell}={\boldsymbol \ell_{+}},$ ${\boldsymbol \ell^{\prime}}=({\boldsymbol \ell_{+}}+{\boldsymbol \ell_{-}})/2.$ 
(See Ref.~\cite{moodley} for details). Here 
\ba
Q(\boldsymbol{\ell},\boldsymbol{\ell}^{\prime})=
{\ell^2\over 4}{\cal N}_{\ell}{1\over 2}{
{\boldsymbol{\ell}\cdot\boldsymbol{\ell}^{\prime}C_{\ell^{\prime}}
+\boldsymbol{\ell}\cdot(\boldsymbol{\ell}-\boldsymbol{\ell}^{\prime})C_{\vert\boldsymbol{\ell}^{\prime}-\boldsymbol{\ell}\vert}} \over
{\big[\tilde C_{\ell^{\prime}}+N_{\ell^{\prime}}\big]
\big[\tilde C_{\vert\boldsymbol{\ell}-\boldsymbol{\ell}^{\prime}\vert}+N_{\vert\boldsymbol{\ell}-\boldsymbol{\ell}^{\prime}\vert}\big]}
}
\ea
where $N_{\ell}$ is the noise power spectrum and $\cal{N}_{\ell}$ is the variance of the estimator defined by
\ba
{\cal N}_{\ell}=\int{d^2{\boldsymbol \ell}^{\prime}\over (2\pi)^2}{1\over 2}{
{\big[\boldsymbol{\ell}\cdot\boldsymbol{\ell}^{\prime}C_{\ell^{\prime}}
+\boldsymbol{\ell}\cdot(\boldsymbol{\ell}-\boldsymbol{\ell}^{\prime})C_{\vert\boldsymbol{\ell}^{\prime}-\boldsymbol{\ell}\vert}\big]^2}
\over
{\big[\tilde C_{\ell^{\prime}}+N_{\ell^{\prime}}\big]\big[\tilde C_{\vert\boldsymbol{\ell}-\boldsymbol{\ell}^{\prime}\vert}+N_{\vert\boldsymbol{\ell}-\boldsymbol{\ell}^{\prime}\vert}\big]}
}.
\label{eqn:var_est_def}
\ea
In the $\ell$ integration we sum modes $2\leq\ell,\ell^{\prime}\leq1200,$ which comprise the modes observable by WMAP.
The validity range of the real--space estimator is defined by the largest and smallest lensing modes that can be probed by the kernel. Hence, the lower $\ell$ limit is determined by the size of the maps, whereas the upper $\ell$ limit is determined by the size of the kernel \cite{moodley}. 

\section{Data and Simulations}
\label{sec:wmap_data}


The relevant information for the reconstruction of the lensing potential lies on the angular scales close to the resolution scale of the map, which is set by the beam size. Hence we opt to use the maps produced by the four W--band differential arrays (DA) for having the smallest beam size\footnote{http://lambda.gsfc.nasa.gov/product/map/dr4/m\_products.cfm}. To minimize the contamination of the lensing signal by spurious sources of non--Gaussianity, we use the coadded, foreground--removed temperature maps \cite{david} 
at 
resolution r9\footnote{http://lambda.gsfc.nasa.gov/product/map/dr4/maps\_da\_r9\_iqu\_7yr\_get.cfm}.
Moreover, to exclude foreground--contaminated portions of the sky at  the location of the galactic plane and point sources, we use the pixel mask KQ85 which keeps 78\% of the sky area\footnote{http://lambda.gsfc.nasa.gov/product/map/dr4/ptsrc\_catalog\_get.cfm}.


We conceive different experiments based on the specifications of the four DAs of the WMAP W--band\footnote{http://lambda.gsfc.nasa.gov/product/map/dr4/beam\_xfer\_get.cfm}. 
For each differential array, we conceive two experiments: a) the experiment denoted by ``theo," which consists of lensed CMB maps synthesised in the flat--sky approximation using the temperature (TT) and lensing deflection ($\alpha\alpha$) power spectra computed with CAMB \cite{camb} (on scales $\ell\leq1200,$ for a fiducial flat $\Lambda$CDM cosmology consisting of $h=0.7,$ $\Omega_{\rm cdm}h^2=0.122,$ $\Omega_{\rm b}h^2=0.0226$, $n_{\rm s}=0.96$, $\sigma_{8}=0.85,$ $\tau_{\rm rei}=0.09$ and $T_{0}=2.725~K$), and b) the experiment denoted by ``obs," which consists of temperature maps derived from HEALPix\footnote{http://healpix.jpl.nasa.gov/} \cite{healpix} cartesian projections of the full--sky map produced by WMAP. 
The specifications of the experiments are summarized in Table \ref{table:expt}.

\begin{table}[t]
\begin{tabular}{c|cccccc}
\hline
Experiment
~&~$\theta_{\rm fwhm}$~&~$\sigma_{0}$~&~$\big<N_{\rm obs}\big>_{\rm pix}$ 
&${\rm FOV_{map}}$~&~$\theta_{\rm kernel}$~\\ 
~&~(arcmin)~&~$(\mu K/N_{\rm obs})$& 
&(deg)~&~(deg)  
\\ \hline
${\rm WMAP}_{W_{1}^{\rm theo/obs}}$~~& 13.2 &$5.906\times 10^3 $ &$2.512\times10^3$ &56 &1.1\\ 
${\rm WMAP}_{W_{2}^{\rm theo/obs}}$~~& 13.2 &$6.572\times 10^3 $ &$2.515\times10^3$ &56 & 1.1\\ 
${\rm WMAP}_{W_{3}^{\rm theo/obs}}$~~& 13.2 &$6.941\times 10^3 $ &$2.514\times10^3$ &56 & 1.1\\ 
${\rm WMAP}_{W_{4}^{\rm theo/obs}}$~~& 13.2 &$6.778\times 10^3 $ &$2.517\times10^3$ &56 &1.1\\ 
\hline
\end{tabular}
\caption{\baselineskip=0.5cm{
{\bf The specifications of the experiments used in this study.} 
The experiments denoted by ${\rm WMAP}_{W_{i}}^{\rm theo}$ use the power spectra computed with CAMB and detector noise characteristic of the $W_{i}$ DA. The experiments denoted by ${\rm WMAP}_{W_{i}}^{\rm obs}$ use cartesian projections of the corresponding $W_{i}$ DA measured by WMAP. }
}
\label{table:expt}
\end{table}

\section{Lensing from WMAP--like simulations}
\label{sec:wmap_ps}

\subsection{The input maps}

We first reconstruct the convergence map from simulations of the four DAs of the WMAP W--band. We synthesise lensed CMB temperature maps from the theoretical power spectra (TT and $\alpha\alpha$) computed with CAMB by shifting the temperature maps by the lensing deflection maps, according to the description in the Appendix of Ref.~\cite{lewis05}. We then combine the lensed temperature maps with the beam window function and detector noise characteristic of each DA of the WMAP W--band, according to the description in Appendix B of Ref.~\cite{moodley}. 
We compute the noise power spectrum from a full--sky map synthesised as ${\textrm{White Noise}}\times\sigma_{0}/\sqrt{N_{\rm obs}},$ where $\sigma_{0}$ is the rms of the noise per number of observations and $N_{\rm obs}$ is the corresponding map containing the number of observations per pixel. From the noise power spectrum, we synthesise a noise map which we add to the beam--convolved, lensed CMB map. 
We call these experiments ${\rm WMAP}^{\rm theo}_{W_{i}}.$ 

For each experiment, we also generate unlensed CMB temperature maps from the theoretical TT power spectrum computed with CAMB. The unlensed CMB maps are generated using different realizations of the CMB map for the same cosmology, thus simulating different realizations of the temperature and noise maps. The unlensed CMB maps serve to model the Gaussian contribution of the CMB to the four--point correlations measured by the estimator of the lensing power spectrum which is the major contaminant of the lensing signal \cite{kesden03}. 

\subsection{The convergence maps}

The convergence map $\hat\kappa(\boldsymbol\theta)$ is estimated by convolving in real space the minimum variance kernel $W(\boldsymbol\theta,\boldsymbol\theta_{+},\boldsymbol\theta_{-})$ with the square of the temperature map from which we want to estimate the weak lensing signal. Both the lensed CMB power spectrum and a fiducial power spectrum for the unlensed CMB power spectrum enter in the calculation of the kernel. We generate the kernel for each DA. In the calculation of the kernel we use the lensed power spectrum computed with CAMB combined with the window beam function and the detector noise characteristic of each DA.

\begin{figure}[t]
\setlength{\unitlength}{1cm}
\vskip-1.5cm
\centerline{
\includegraphics[width=10cm]
{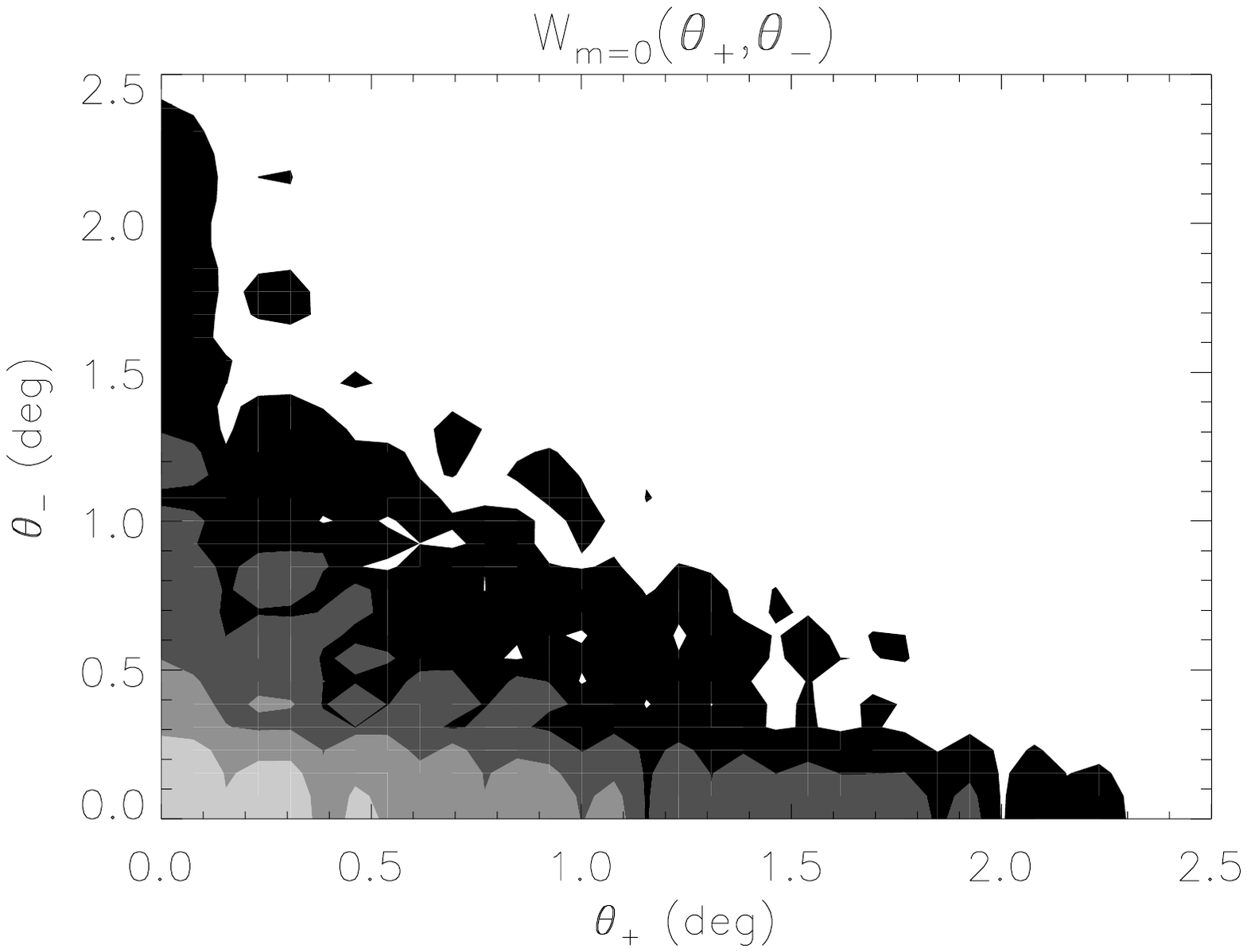}
\hskip-2cm
\includegraphics[width=10cm]
{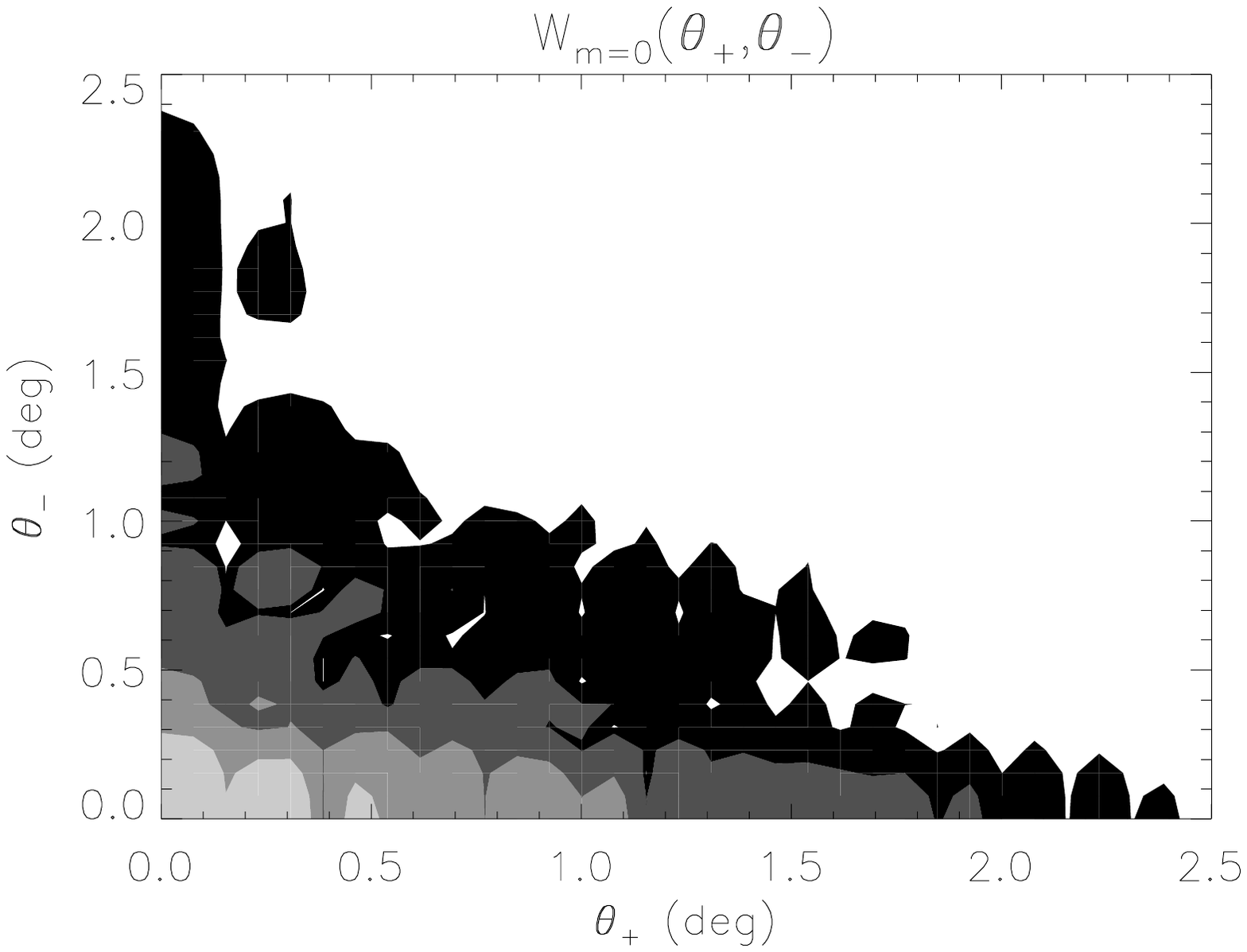}
}
\vskip-7cm
\centerline{
\includegraphics[width=10cm]
{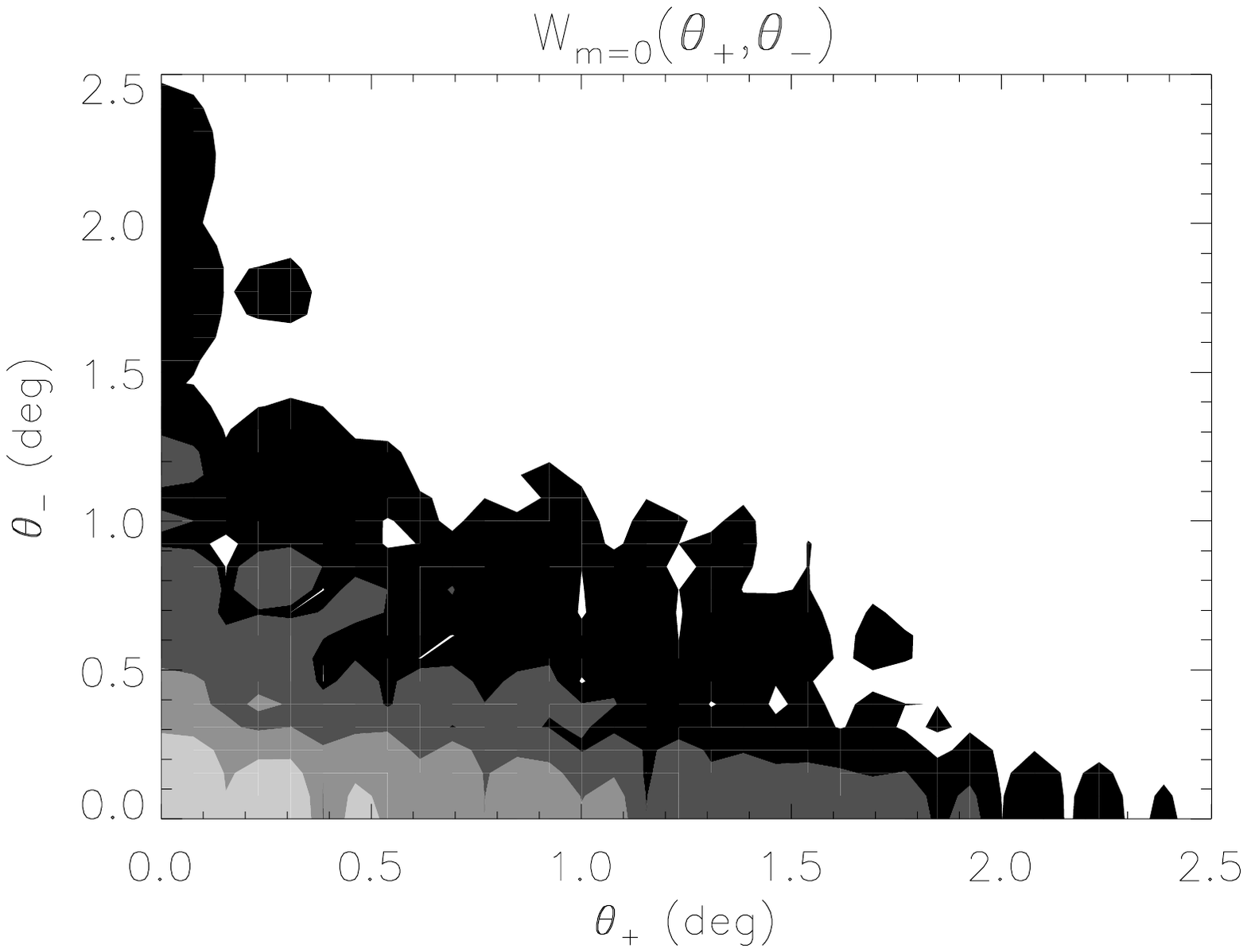}
\hskip-2cm
\includegraphics[width=10cm]
{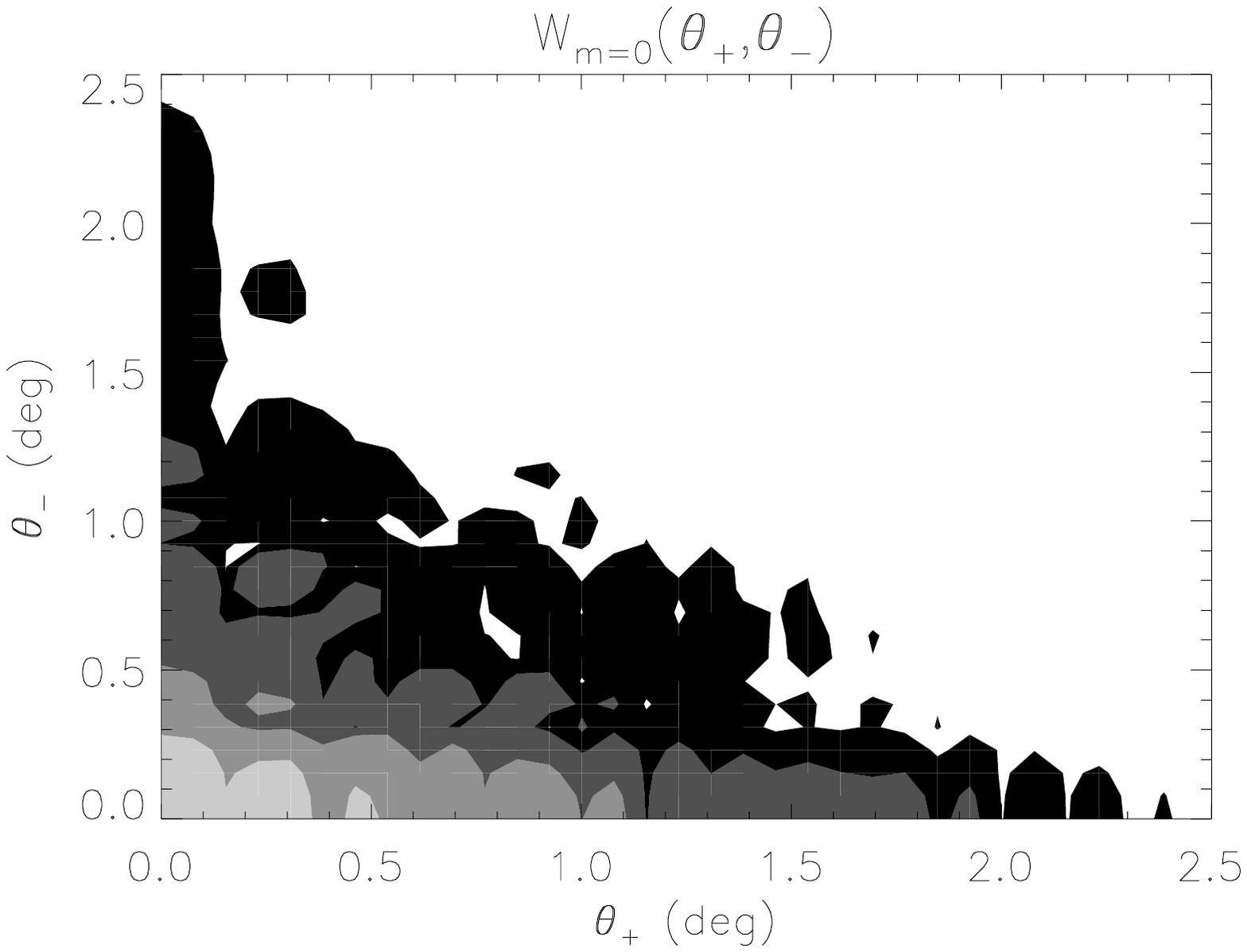}
}
\vskip-6.cm
\caption{\baselineskip=0.5cm{ 
{\bf Kernel in real space for the four DA's of the WMAP W--band (theo).}
We plot the dominant ($m=0$) eigenfunction only. The levels denote the orders of magnitude as fractions of the maximum value, delimited by 
$\{1,10^{-1},10^{-2},10^{-3},10^{-4}\}.$
The kernels in all panels are derived from the theoretical power spectra combined with the beam window function of each DA. 
Top left: $W_1,$ Top right: $W_2,$ Bottom left: $W_3,$ Bottom right: $W_4.$
}}
\label{fig:kernel_designer_da}
\end{figure}

The kernel can be expressed as a series of eigenfunctions, the dominant being the $m=0$ mode which we plot in Fig.~\ref{fig:kernel_designer_da}. (See Ref.~\cite{moodley} for the calculation of the kernel in real space.) The plots show that the kernels for the four experiments have the same size and similar structures within the  $1\%$ level, differing only in the smaller--scale structures (of order $<0.1^{\circ}$) at sub--percent level, so we expect that the corresponding lensing reconstructions have similar ranges of validity. In all cases, for the numerical implementation we truncate the kernel at $\theta_{\rm kernel}=1.1^{\circ},$ which is the spatial extent which corresponds to the $1\%$ level. As previously tested, keeping sub $1\%$ levels does not improve the reconstruction of the power spectrum \cite{moodley}.

The convergence map $\hat \kappa$ reconstructed from a lensed CMB map contains both the lensing signal $\kappa_{\vert\psi}$ and the contribution  $\kappa_{\vert\psi=0}$ due to the unlensed CMB map, which we write schematically as $\hat \kappa=\kappa_{\vert\psi}+\kappa_{\vert\psi=0}.$ The zero--point function of this contribution amounts to an offset which we remove by subtracting the pixel average of the reconstructed convergence map $\hat \kappa.$ The resulting convergence map $\hat \kappa-\big<\hat\kappa\big>_{\text{pixel}}$ is unbiased  in the sense that its expectation value is equal to that of  $\kappa_{\vert\psi}.$ (See Refs.~\cite{moodley,tereno} for details.) 

\subsection{The convergence power spectrum}

The convergence power spectrum is computed directly from the unbiased convergence map. Its variance can be computed analytically from the variance of the convergence estimator ${\rm Var}[\hat \kappa]$ given by Eqn.~(\ref{eqn:var_est_def}), while including also the cosmic variance, yielding
\ba
{\rm Var}\big[C_{\ell}^{\hat \kappa}\big]
={1\over {\ell f_{\rm sky}}}\left[
{\rm Var}[\hat \kappa]
+C_{\ell}^{\hat \kappa}\right]^2.
\label{eqn:var_est0}
\ea
(See e.g. Ref.~\cite{hu01} for a derivation.)
In addition to the lensing signal, the convergence power spectrum also includes a Gaussian contribution due to the unlensed CMB (the diagonal correlations mentioned in Sec.~ \ref{sec:wmap_begin}). 

Following Ref.~\cite{tereno}, we compute the Gaussian contribution using an ensemble of unlensed CMB maps synthesised from a common CAMB temperature power spectrum by varying the seed of each realization. A convergence map is reconstructed from each realization of the unlensed CMB and the corresponding convergence power spectrum is computed. The weighted mean of these power spectra is the estimate of the Gaussian bias that we must subtract off of the power spectrum of the convergence map reconstructed from the lensed CMB map. It is given by
\ba
\big<C_{\ell}^{\hat \kappa_{\vert\psi=0}}\big>_{CMB}
={\sum_{i}^{\textrm{num maps}}
\left(C_{\ell}^{\hat \kappa_{\vert\psi=0}}\right)_{i}~w_{i}^{\hat\kappa_{\vert\psi=0}}}
/{\sum_{i}^{\textrm{num maps}}w_{i}^{\hat\kappa_{\vert\psi=0}}}.
\label{eqn:cl_unlensed_w}
\ea
The weights $w_{i}^{\hat\kappa_{\vert\psi=0}}$ are defined as the inverse of the variance of each $\big(C_{\ell}^{\hat \kappa_{\vert\psi=0}}\big)_{i}$
\ba
{\rm Var}\big[C_{\ell}^{\hat \kappa_{\vert\psi=0}}\big]_{i}
={1\over {\ell f_{\rm sky}}}\left[
{\rm Var}[\hat \kappa]
+\left(C_{\ell}^{\hat \kappa_{\vert\psi=0}}\right)_{i}\right]^2.
\label{eqn:var_est}
\ea
The weights do not differ much, differing only by the cosmic variance contribution.
The power spectrum corrected for the Gaussian bias is given by
\ba
C_{\ell}^{\hat\kappa}
-\big<C_{\ell}^{\hat \kappa_{\vert\psi=0}}\big>_{CMB},
\label{eqn:oldestimator}
\ea
which is a difference between two very similar quantities. The variance of this quantity is given by
\ba
{\rm Var}\big[C_{\ell}^{\hat \kappa}\big]
+{\rm Var}\big[\big<C_{\ell}^{\hat \kappa_{\vert\psi=0}}\big>_{CMB}\big],
\label{eqn:var_cl_lens_w}
\ea
where we include the uncertainty on the Gaussian bias which is given by the variance of the mean of the ensemble of the realizations, i.e.,
\ba
{\rm Var}\big[\big<C_{\ell}^{\hat \kappa_{\vert\psi=0}}\big>_{CMB}\big]
=\left[\sum_{i}^{\textrm{num maps}}\left(1/{\rm Var}\big[C_{\ell}^{\hat \kappa_{\vert\psi=0}}\big]_{i}\right)\right]^{-1}.
\label{eqn:var_cl_unlensed_w}
\ea

\begin{figure}[t]
\setlength{\unitlength}{1cm}
\centerline{
\hskip-0.5cm
\includegraphics[width=12cm]
{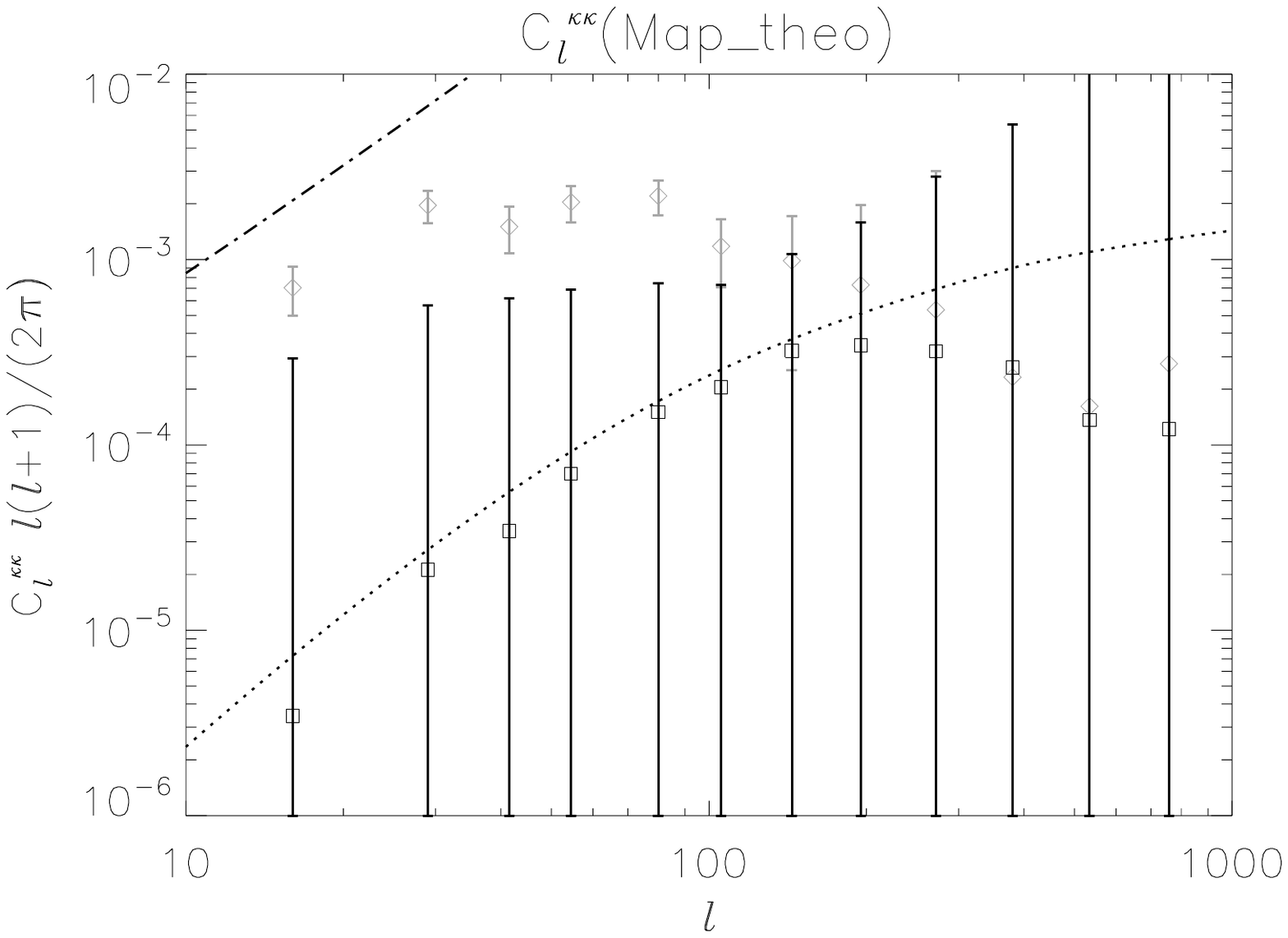}
\hskip-3cm
\includegraphics[width=12cm]
{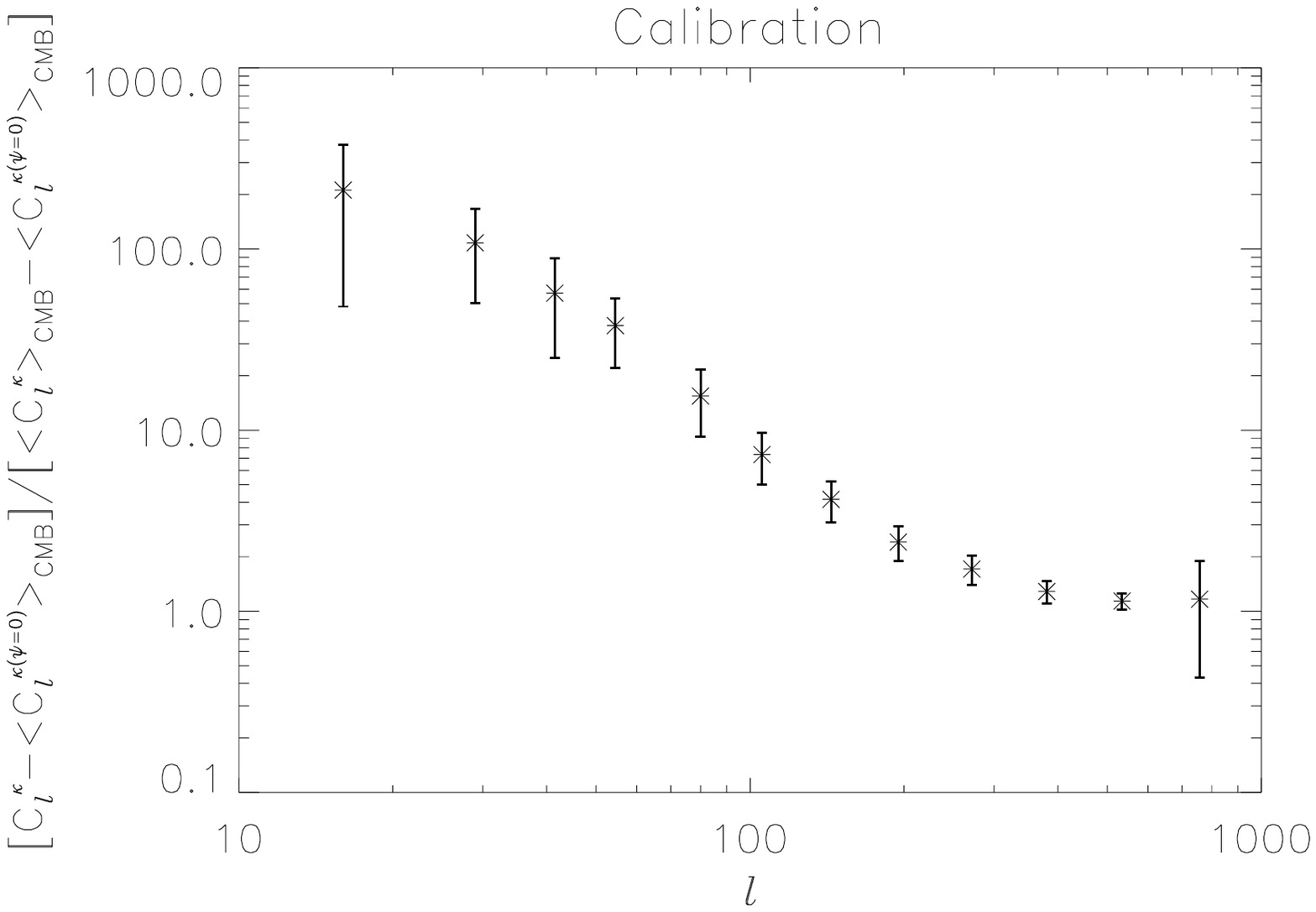}
}
\vskip-7.5cm
\caption{\baselineskip=0.5cm{ 
{\bf 
Convergence power spectrum computed from CMB simulations of the WMAP's W--band DAs.}
Left panel: The gray diamonds are the convergence power spectrum computed as the difference between one lensed CMB map and the inverse--variance average over different realizations of unlensed CMB maps. 
The black squares are the convergence power spectrum computed as the difference between the inverse--variance average over different realizations of paired lensed CMB and unlensed CMB maps. For each DA, the maps are synthesised from the theoretical power spectra combined with the beam window function and noise power spectrum of the corresponding DA. The dotted line is the theoretical convergence power spectrum computed with CAMB. The dash--dotted line is the inverse--sum of the variance of the estimator ${\rm Var}[\hat \kappa_{\vert\psi}]$ for each DA.
Right panel: The black stars show the bias calibration $f_{\rm cal}.$ 
}}
\label{fig:cl_kk2_designer_da}
\end{figure}

For each of the four experiments ${\rm WMAP}^{\rm theo}_{W_{i}},$ we obtain a convergence power spectrum using Eqn.~(\ref{eqn:oldestimator}). The average is plotted in Fig.~\ref{fig:cl_kk2_designer_da} (left panel, gray diamonds). The bias uncertainty becomes subdominant in the entire $\ell$ range with the increase of the number of maps. The detection error is dominated by the estimator noise set by the first term in the right--hand side of  Eqn.~(\ref{eqn:var_est0}), which depends on the experimental resolution and detector noise. The decrease of power for large $\ell$s marks the end of the validity range of the real--space estimator.

However, we observe that the result contains a large excess of power as compared to the input convergence power spectrum, also shown in Fig.~\ref{fig:cl_kk2_designer_da}. This is in contrast with the results in Refs.~\cite{moodley, tereno}, where the same method applied to simulations of the Planck experiment was able to retrieve the input signal. This indicates that in the comparatively lower signal--to--noise WMAP experiment, the quantity $\big<C_{\ell}^{\hat \kappa_{\vert\psi=0}}\big>_{CMB}$ is no longer a good estimator of all non--negligible biases.  A source of additional non--Gaussian bias are couplings of lensing modes that derive from higher--order terms in $\psi$ in the Taylor expansion and that also contribute to the four--point correlation function of the lensed CMB, which is essentially what the estimated power spectrum measures. We expect to get a much better agreement with the input lensing signal if we are able to reduce the uncertainty in the CMB map, since these couplings add incoherently \cite{kesden03}. 

We then proceed to synthesise an ensemble of lensed CMB maps, all from the same CAMB temperature and lensing power spectra used in the single realization described earlier on. The various realizations use different realizations of the noise map.
A convergence map is reconstructed from each realization of the lensed CMB and the corresponding convergence power spectrum is computed. The mean of these power spectra will no longer contain the spurious lensing correlations. As before, we also generate unlensed CMB maps, estimate the corresponding convergence maps and average the power spectra over the unlensed realizations to obtain the Gaussian bias. We subtract the mean power spectrum of the unlensed realizations from the mean power spectrum of the lensed realizations, obtaining the unbiased power spectrum of the convergence  $C_{\ell}^{\kappa_{\vert\psi}}$ as
\ba
C_{\ell}^{\kappa_{\vert\psi}}
=\big<C_{\ell}^{\hat \kappa}\big>_{CMB}
-\big<C_{\ell}^{\hat \kappa_{\vert\psi=0}}\big>_{CMB}.
\label{eqn:newestimator}
\ea
Here, the first term on the right--hand side averages out the lensing bias, while the second term models the Gaussian bias. Both terms are computed as weighted averages with the weights given by the inverse of the theoretical variance of each realization given by  Eqns.~(\ref{eqn:var_est0}) and (\ref{eqn:var_est}).
The variance of $C_{\ell}^{\kappa_{\vert\psi}}$ is given by 
\ba
{\rm Var}\big[C_{\ell}^{\kappa_{\vert\psi}}\big]
={\rm Var}\big[\big<C_{\ell}^{\hat \kappa}\big>_{CMB}\big]
+{\rm Var}\big[\big<C_{\ell}^{\hat \kappa_{\vert\psi=0}}\big>_{CMB}\big],
\label{eqn:var_cl_lens_ww}
\ea
where ${\rm Var}\big[\big<C_{\ell}^{\hat\kappa}\big>_{CMB}\big]$ is given by 
\ba
{\rm Var}\big[\big<C_{\ell}^{\hat \kappa}\big>_{CMB}\big]
=\left[\sum_{i}^{\textrm{num maps}}\left(1/{\rm Var}\big[C_{\ell}^{\hat \kappa}\big]_{i}\right)\right]^{-1}
\label{eqn:varmean}
\ea
with ${\rm Var}\big[C_{\ell}^{\hat \kappa}\big]_{i}$ given by Eqn.~(\ref{eqn:var_est0}),
and ${\rm Var}\big[\big<C_{\ell}^{\hat \kappa_{\vert\psi=0}}\big>_{CMB}\big]$ is given  by
Eqn.~(\ref{eqn:var_cl_unlensed_w}).

The result obtained with Eqn.~(\ref{eqn:newestimator}) is shown in Fig.~\ref{fig:cl_kk2_designer_da} (left panel, black squares). We observe that it closely reproduces the input power spectrum within the validity range of the real--space estimator. 
Since Eqn.~(\ref{eqn:newestimator}) is correctly normalized to the true lensing power spectrum, we conclude that averaging the convergence power spectra obtained from different lensed CMB realizations removes the spurious lensing correlations measured with Eqn.~(\ref{eqn:oldestimator}). This is similar to the approach suggested in Sec.~V of Ref.~\cite{hanson11} for the removal of the bias in $\psi^2$ (the $N_{\ell}^{(1)}$ bias).

As a measure of the statistical significance of the detection, we compute the best--fit lensing amplitude parameter $A_L$. 
We fit the computed power spectrum with functions of the form $A_L\,C_{\ell}^{\rm in}$, where $C_{\ell}^{\rm in}$ is the input fiducial convergence power spectrum. The $\chi^2$--minimization yields an analytical expression for $A_L$, which is essentially an inverse--variance weighting of the ratio $C_{\ell}^{\rm out}/C_{\ell}^{\rm in}$ summed over all scales $\ell$ \cite{smith07},
\ba
A_L={\sum_{\ell}C_{\ell}^{\rm in}C_{\ell}^{\rm out}/{\rm Var}[C_{\ell}^{\rm out}]\over
\sum_{\ell} (C_{\ell}^{\rm in})^2/{\rm Var}[C_{\ell}^{\rm out}]}.
\label{eqn:A_L}
\ea
The error of $A_{L}$ is computed by propagating the error of Eqn.~(\ref{eqn:A_L}), yielding 
\ba
{\rm Var}[A_{L}]
={1
\over \sum_{\ell} (C_{\ell}^{\rm in})^2/{\rm Var}[C_{\ell}^{\rm out}]}.
\label{eqn:Var_A_L}
\ea
We compute $A_{L}$ for each DA, $(A_{L}^{\rm theo})_{W_{i}},$ for the average of the four DA's auto--correlations, $(A_{L}^{\rm theo})_{\left<W_{i}W_{i}\right>}$, and for the full combination of DA's considering all auto and cross--correlations, $(A_{L}^{\rm theo})_{\left<W_{i}W_{j}\right>}.$  
The results for the four experiments, separately and combined, in the validity range of the estimator, are presented in Table~\ref{table:A_L_designer}.
For the combined result we find $A_{L}=1.01\pm 1.46,$ which is consistent with the fiducial lensing power spectrum used as input.

\begin{table}[t]
\begin{tabular}{c||cccc|cc}
\hline
~&~$(A_{L}^{\rm theo})_{W_1}$~&~$(A_{L}^{\rm theo})_{W_2}$
~&~$(A_{L}^{\rm theo})_{W_3}$~&~$(A_{L}^{\rm theo})_{W_4}$
~&~$(A_{L}^{\rm theo})_{\left<W_iW_i\right>}$
~&~$(A_{L}^{\rm theo})_{\left<W_iW_j\right>}$
\\ \hline
$\ell \in  [6,198]$
& $0.98\pm4.67$ & $1.01\pm5.96$ & $1.04\pm6.22$ & $1.05\pm5.48$
& $1.02\pm2.74$ & $1.01\pm1.46$
\\ \hline 
\end{tabular}
\caption{\label{table:A_L_designer} \baselineskip=0.5cm{
{\bf $A_L^{\rm theo}$ from each W--band DA 
and from the combined W--band DAs.} 
}}
\end{table}

\section{Lensing from WMAP data}
\label{sec:wmap_maps}

We now turn to the problem of detecting a lensing signal in the WMAP 7--year data. Having only one full--sky ``realization" of the lensed map, we cannot compute the unbiased power spectrum using Eqn.~(\ref{eqn:newestimator}). 
However, we can still follow the suggested approach in Ref.~\cite{hanson11} where the bias acts as a normalization effect and the estimator is seen as a convolution over the true lensing power spectrum. To deconvolve the bias and produce an unbiased estimate correctly normalized to the true lensing power spectrum, we use the simulation results of the previous section. 
For a given realization of $C_{\ell}^{\hat \kappa}$ in Eqn.~(\ref{eqn:oldestimator}), we define the bias calibration $f_{{\rm cal,} i}$ as the ratio between Eqn.~(\ref{eqn:oldestimator}) and  Eqn.~(\ref{eqn:newestimator}). A distribution of calibration functions $f_{{\rm cal,} i}$ may be defined, one for each power spectrum computed by Eqn.~(\ref{eqn:oldestimator}). The average $f_{\rm cal}$ over the different realizations is shown in Fig.~\ref{fig:cl_kk2_designer_da} (right panel). 

\subsection{The input maps}
\label{subsec:wmap_input_maps}

We then proceed to reconstruct the convergence map from the temperature maps produced by the four DA's of WMAP W--band. We use the foreground--reduced maps and apply the KQ85 mask which keeps 78\% of the sky area. We then make cartesian projections  of the full--sky masked maps into patches of the same size ($56^{\circ}\times 56^{\circ}$) as the simulated maps in the previous section.  The projections cover the whole sphere in steps of 45 degrees along the latitude and along the longitude. In a similar way, we also project the mask and the map containing the number of observations per pixel for each DA which we use to generate the noise patches. We call these experiments ${\rm WMAP}^{\rm obs}_{W_{i}}.$ In Ref.~\cite{plaszczynski11}, a similar strategy is adopted to estimate the bispectrum from a full--sky map of a Planck simulation.

For each patch we generate unlensed CMB maps. These maps consist of the same CMB map combined with the DA's beam window function, plus a noise map computed as ${\textrm{White Noise}}\times\sigma_{0}/\sqrt{N_{\rm obs}},$ where $\sigma_{0}$ is the rms of the noise per number of observations and $N_{\rm obs}$ is the corresponding patch containing the number of observations per pixel. Finally we apply the corresponding mask patch.

\subsection{The convergence maps}

\begin{figure}[t]
\setlength{\unitlength}{1cm}
\vskip-1.5cm
\centerline{
\includegraphics[width=10cm]
{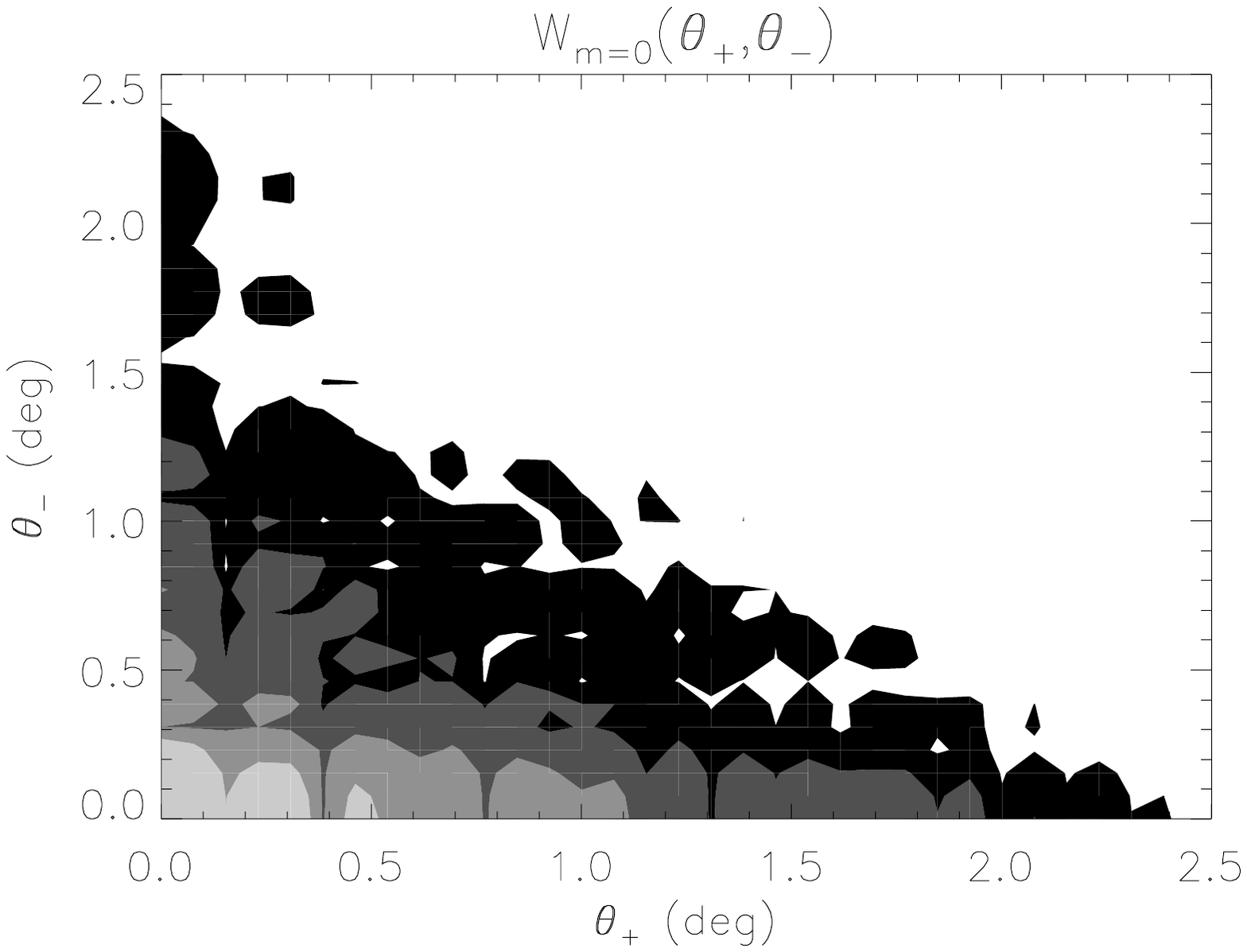}
\hskip-2cm
\includegraphics[width=10cm]
{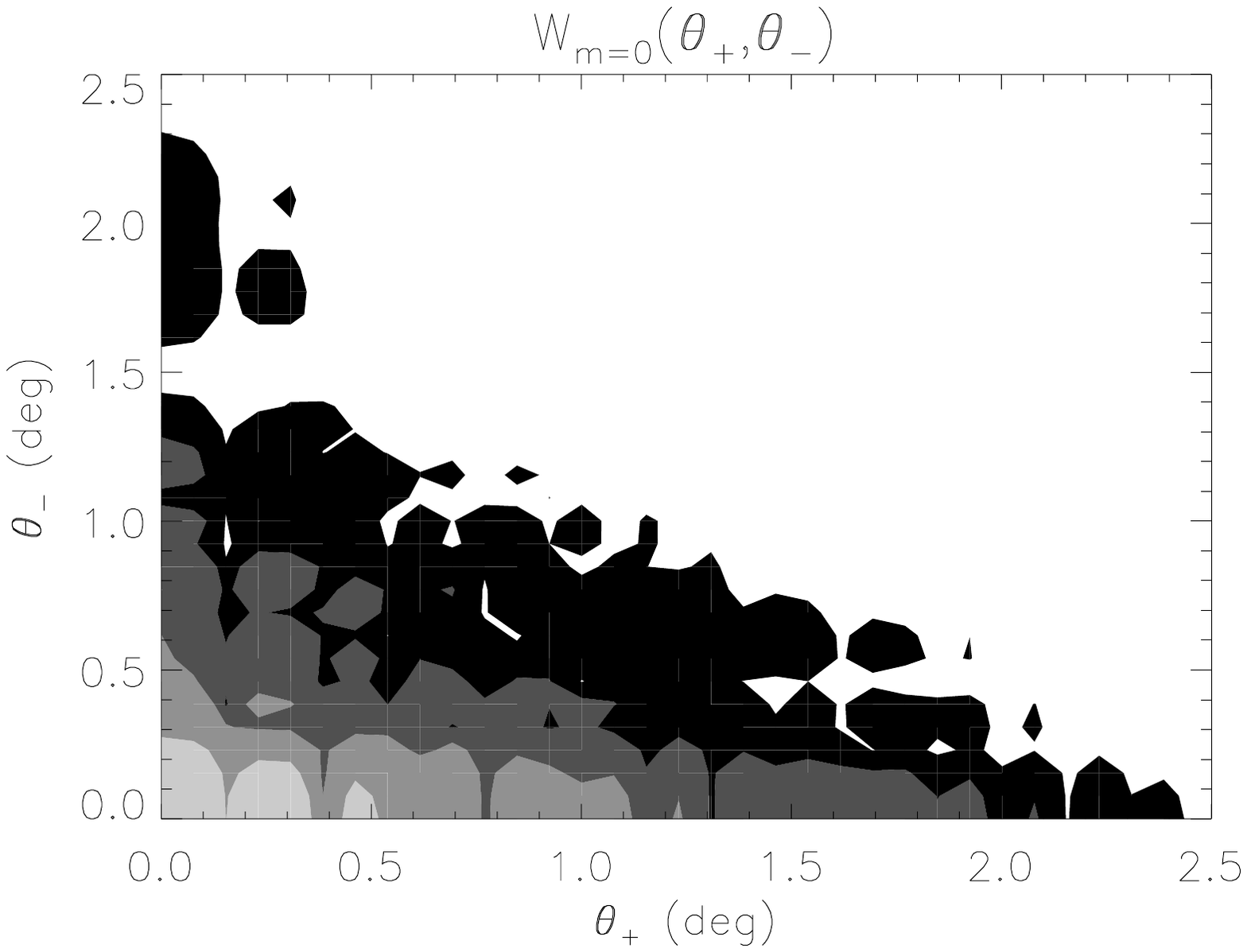}
}
\vskip-7.cm
\centerline{
\includegraphics[width=10cm]
{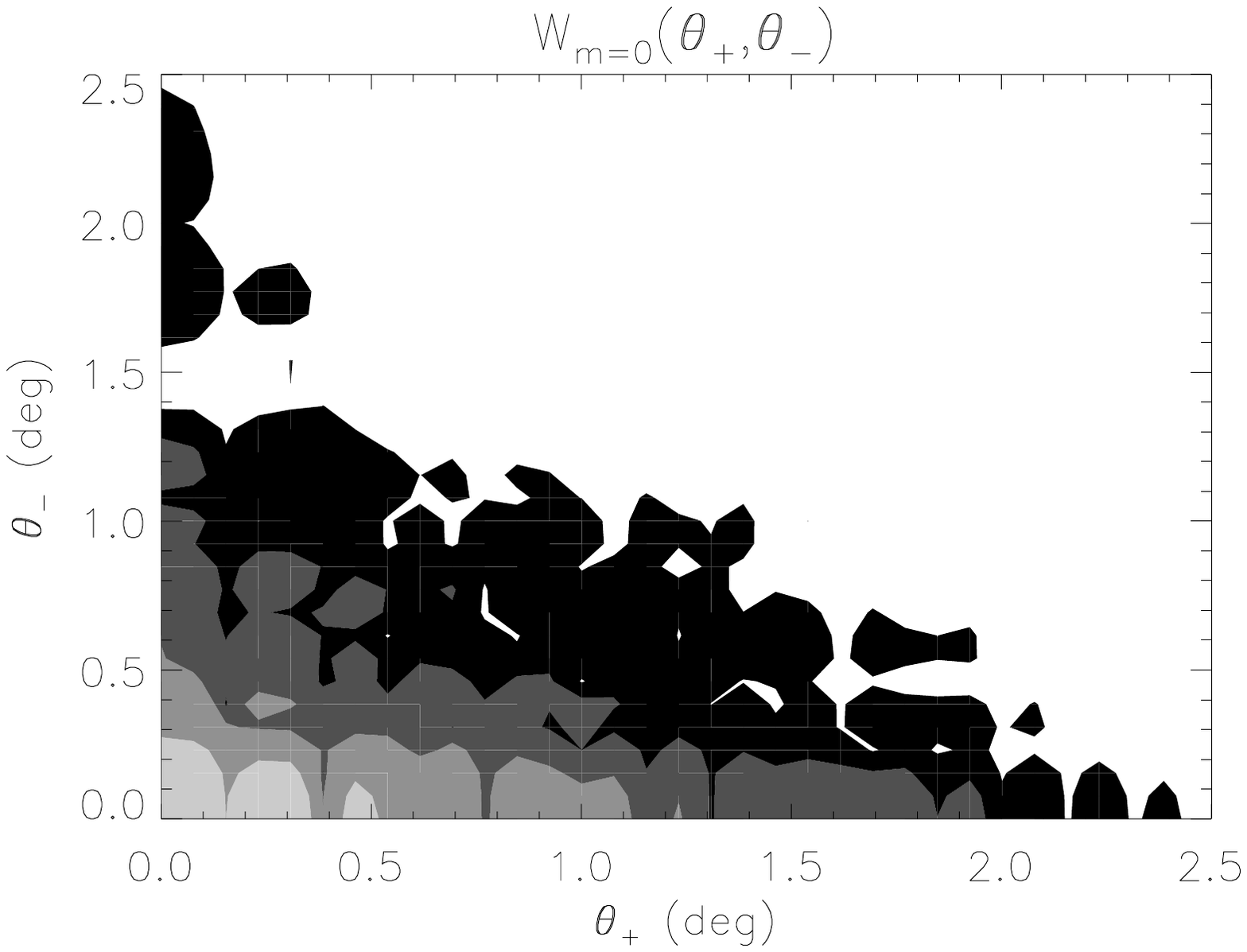}
\hskip-2cm
\includegraphics[width=10cm]
{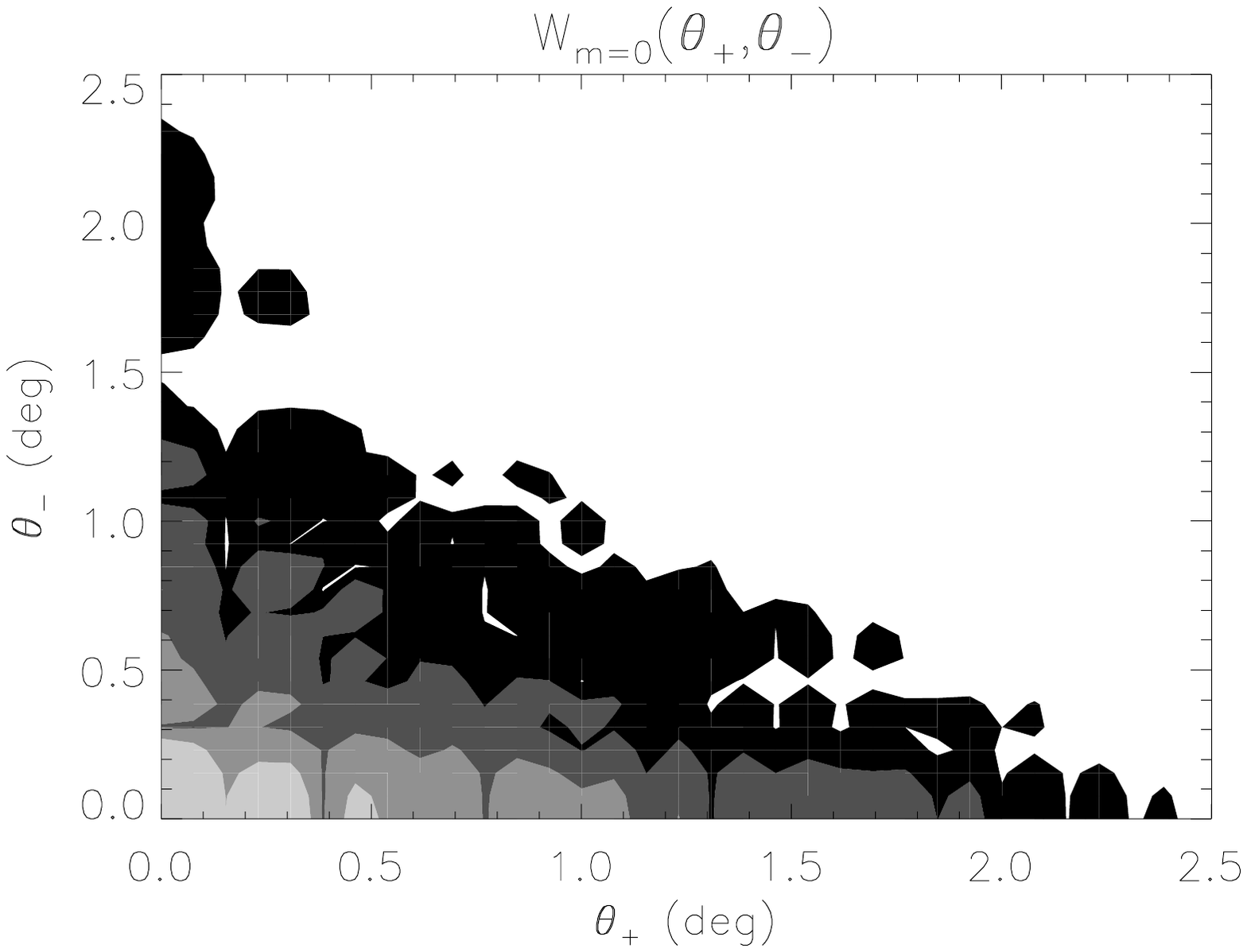}
}
\vskip-6.cm
\caption{\baselineskip=0.5cm{ 
{\bf Kernel in real space for the four DA's of the WMAP W--band (obs).
} We plot the dominant ($m=0$) eigenfunction only. The levels denote the orders of magnitude as fractions of the maximum value, delimited by 
$\{1,10^{-1},10^{-2},10^{-3},10^{-4}\}.$
The kernels in all panels are derived from the WMAP TT power spectrum combined with the beam window function of each DA.
Top left: $W_1,$ Top right: $W_2,$ Bottom left: $W_3,$ Bottom right: $W_4.$
}}
\label{fig:kernel_da}
\end{figure}

\begin{figure}[t]
\setlength{\unitlength}{1cm}
\vskip-1.5cm
\centerline{
\hskip2cm
\includegraphics[width=8cm]
{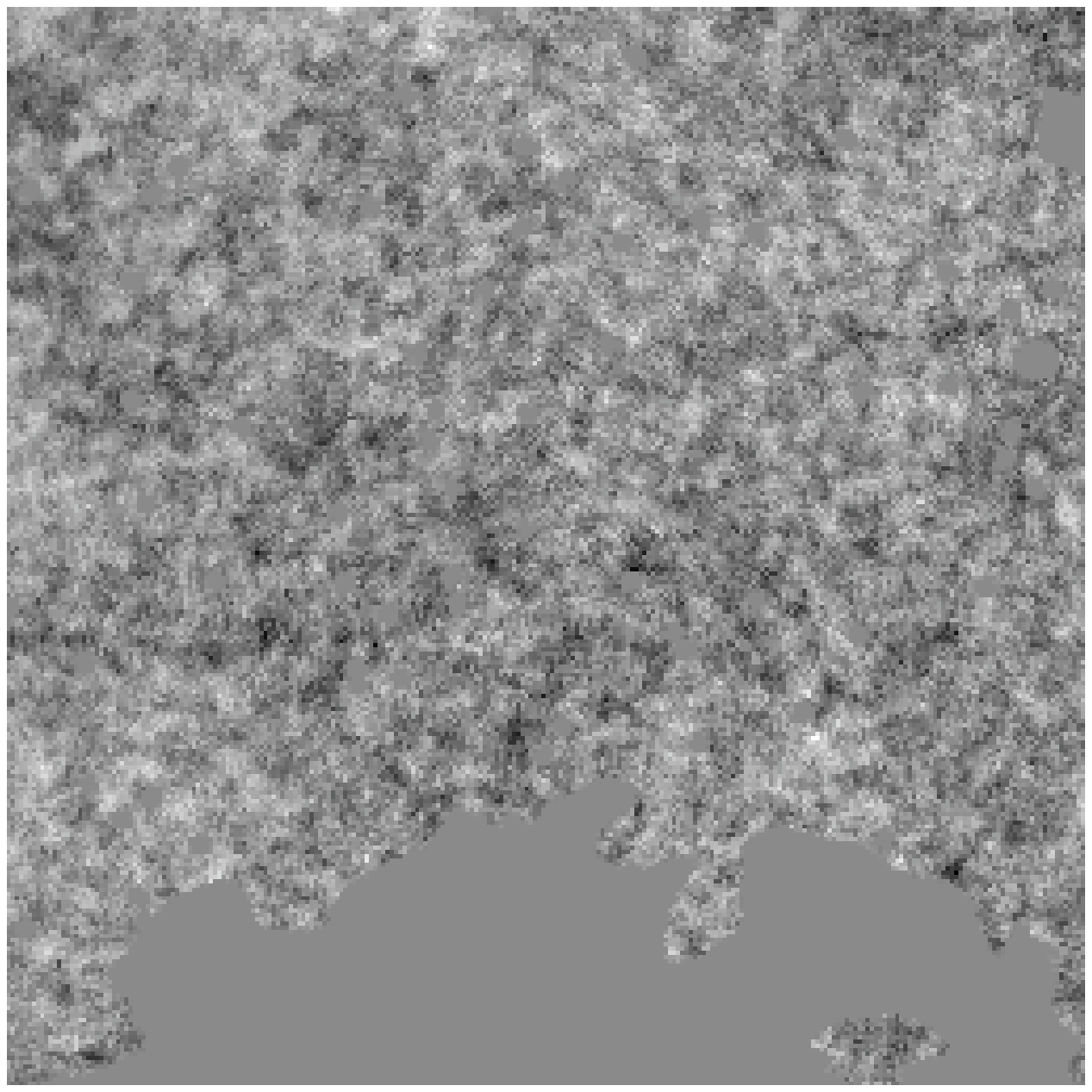}
\hskip-3.cm
\includegraphics[width=8cm]
{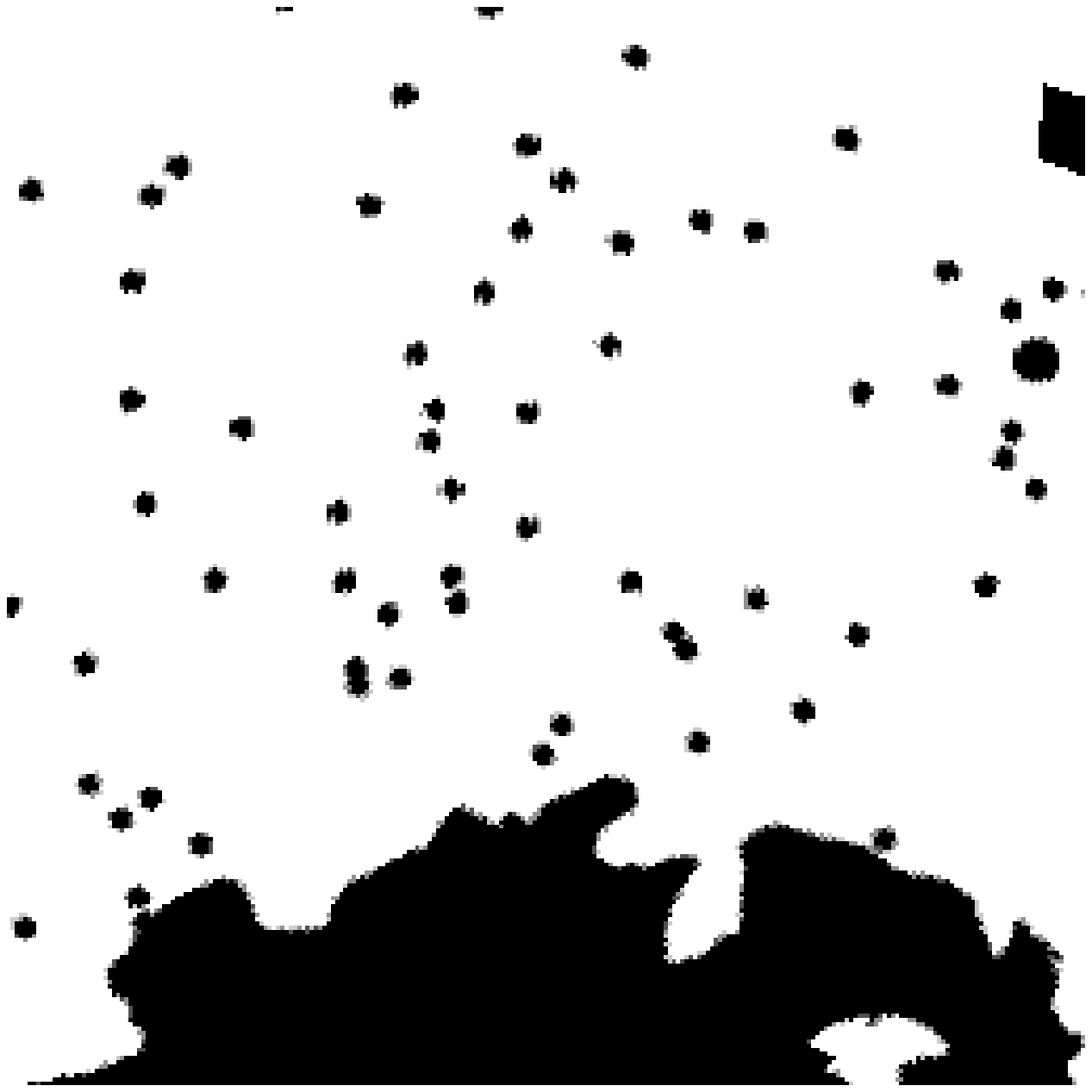}
\hskip-3cm
\includegraphics[width=8cm]
{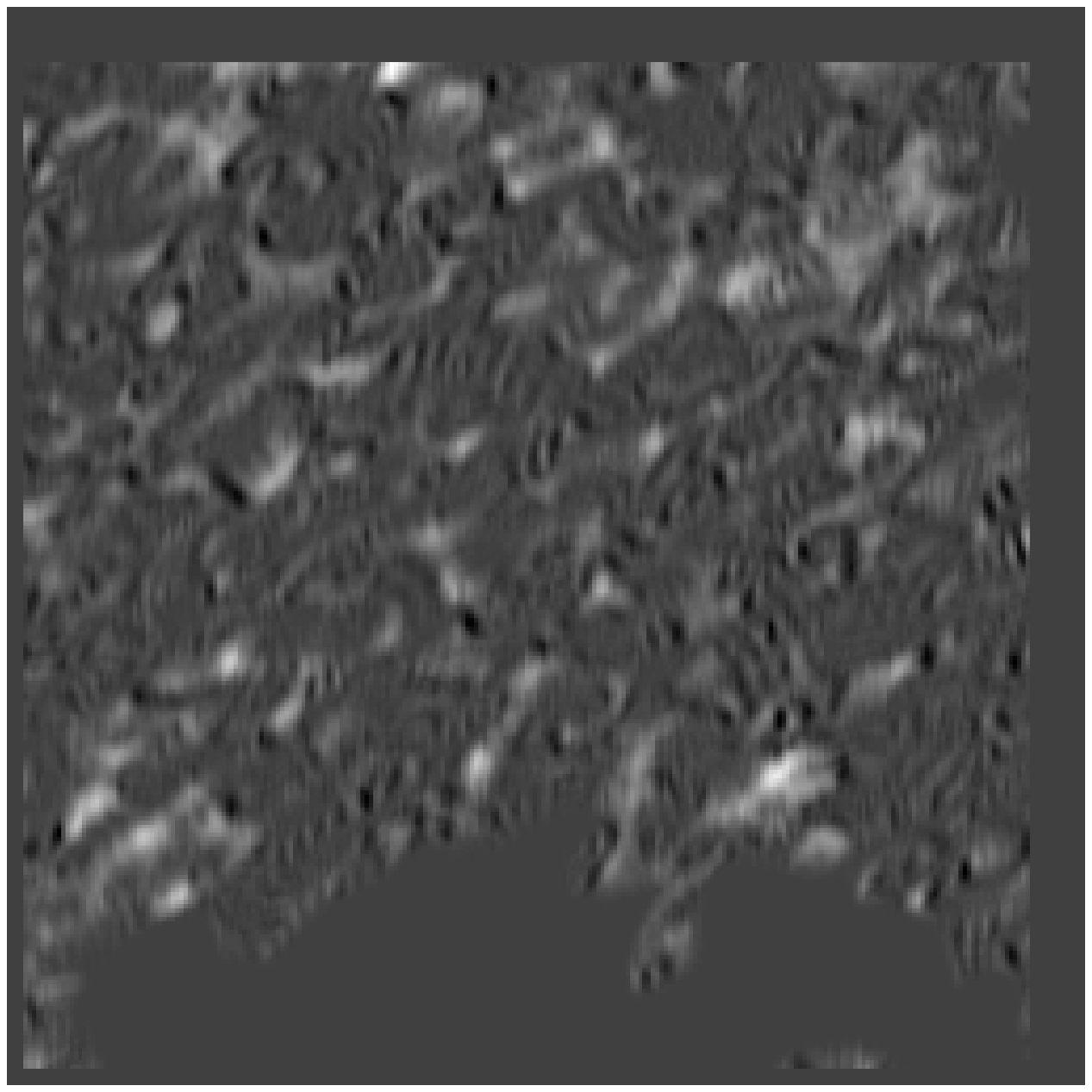}
}
\vskip-5.5cm
\centerline{
\hskip2cm
\includegraphics[width=8cm]
{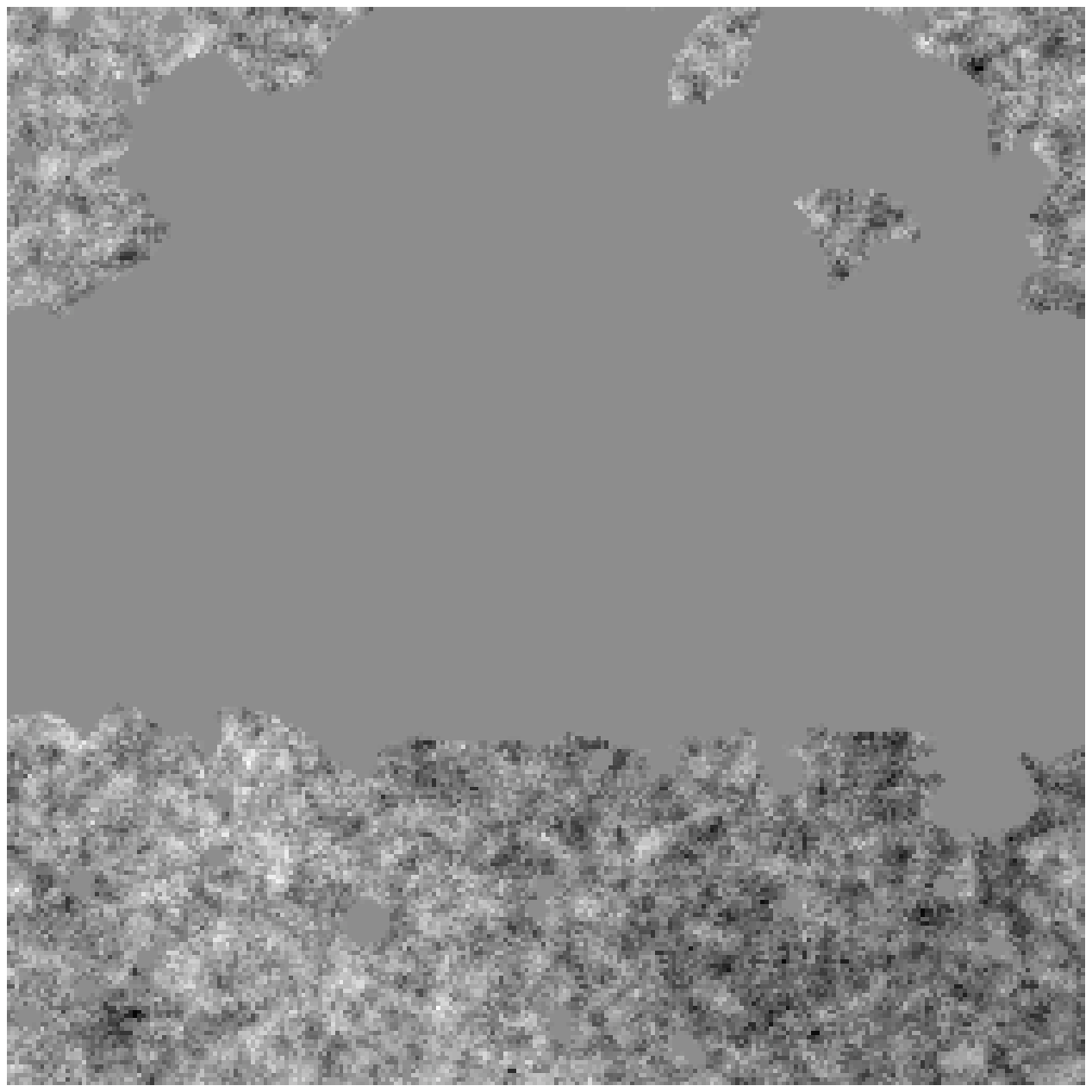}
\hskip-3.cm
\includegraphics[width=8cm]
{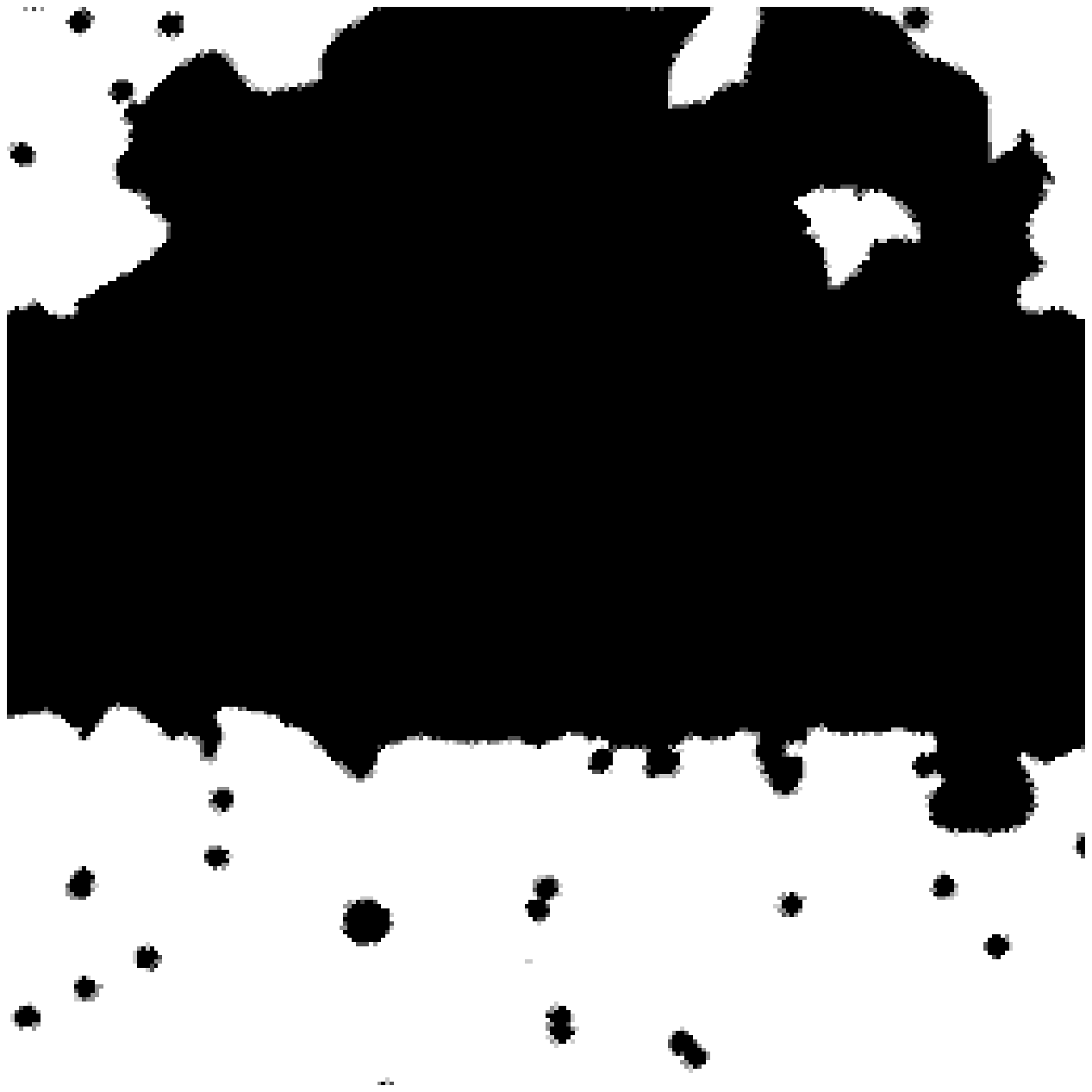}
\hskip-3.cm
\includegraphics[width=8cm]
{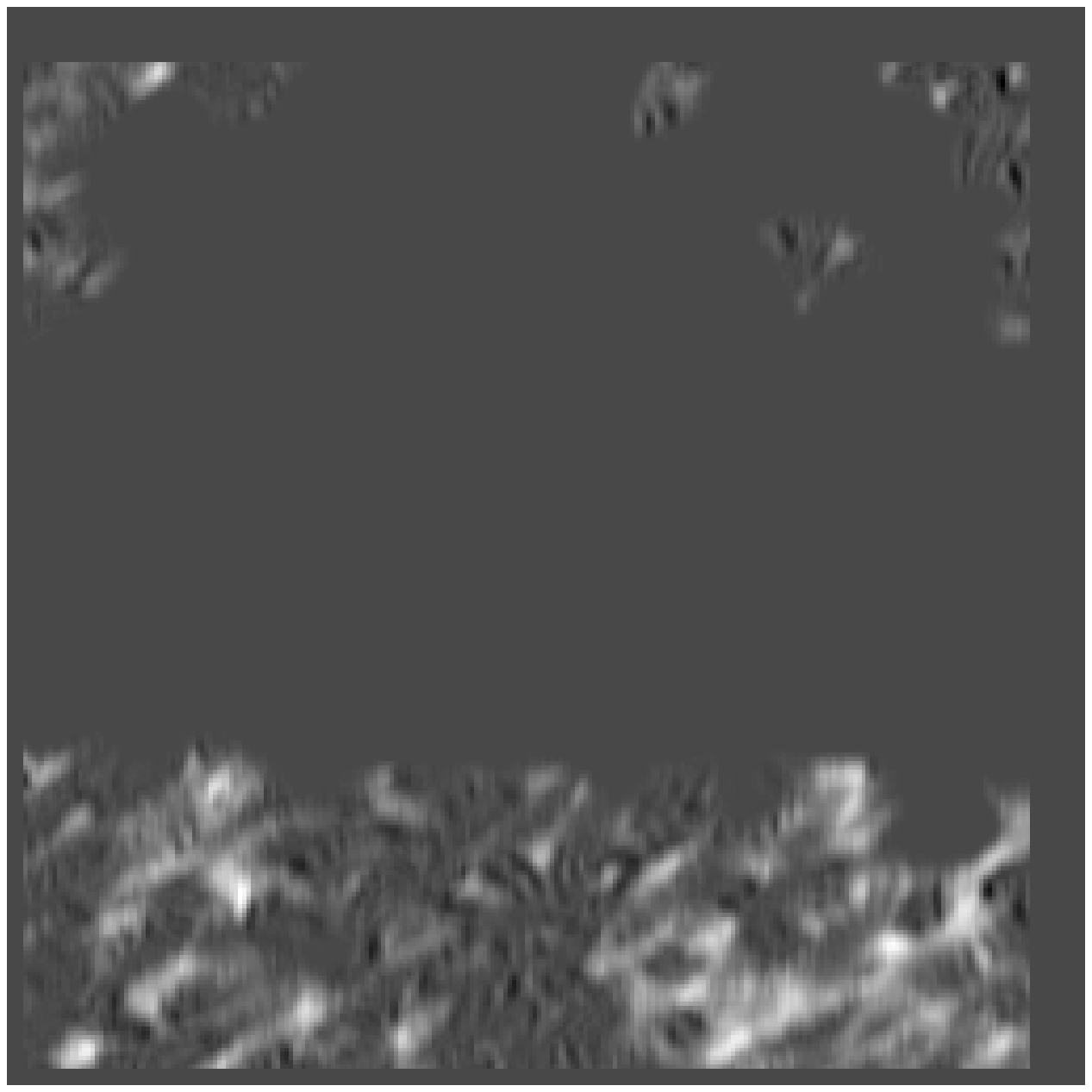}
}
\vskip-5.5cm
\centerline{
\hskip2cm
\includegraphics[width=8cm]
{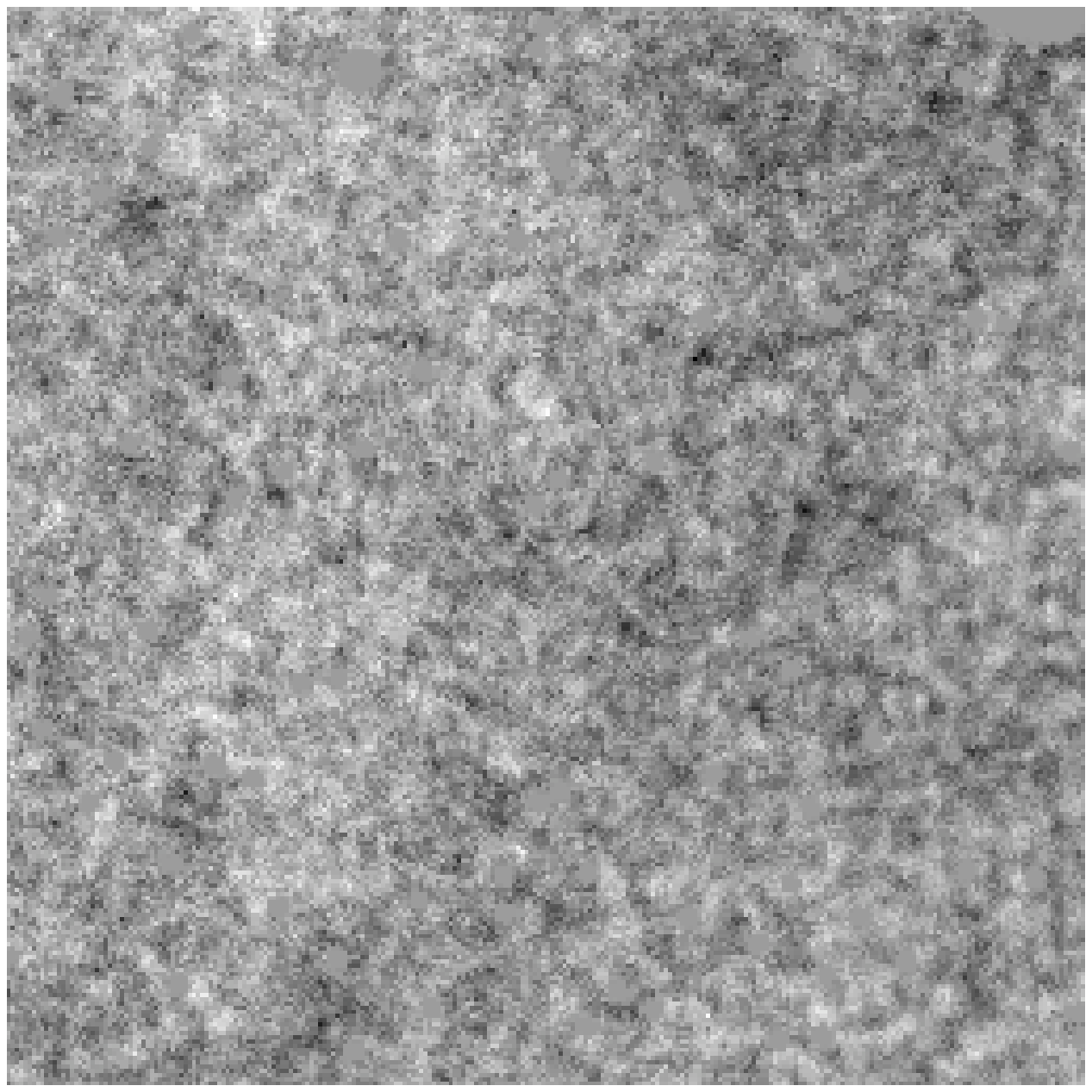}
\hskip-3.cm
\includegraphics[width=8cm]
{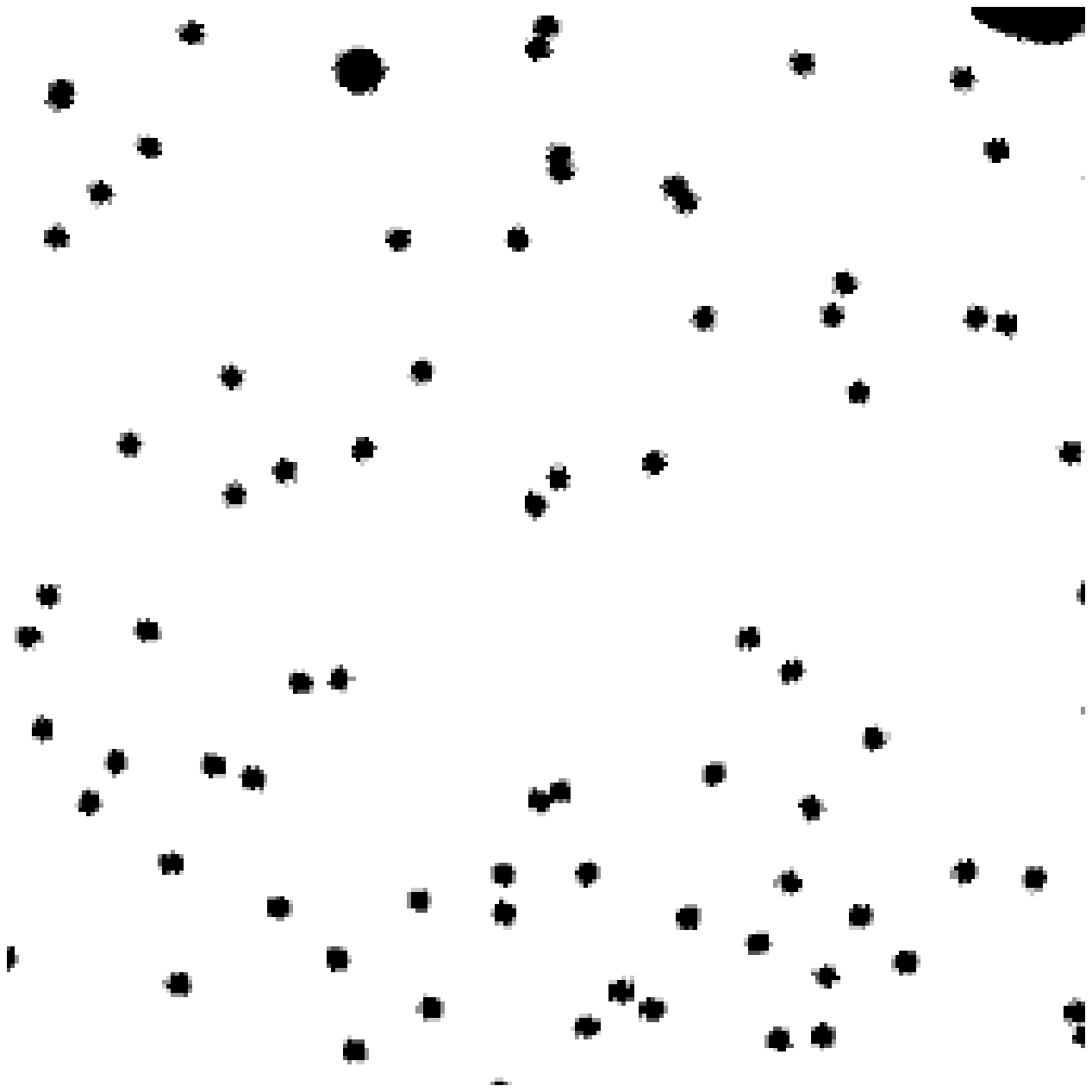}
\hskip-3.cm
\includegraphics[width=8cm]
{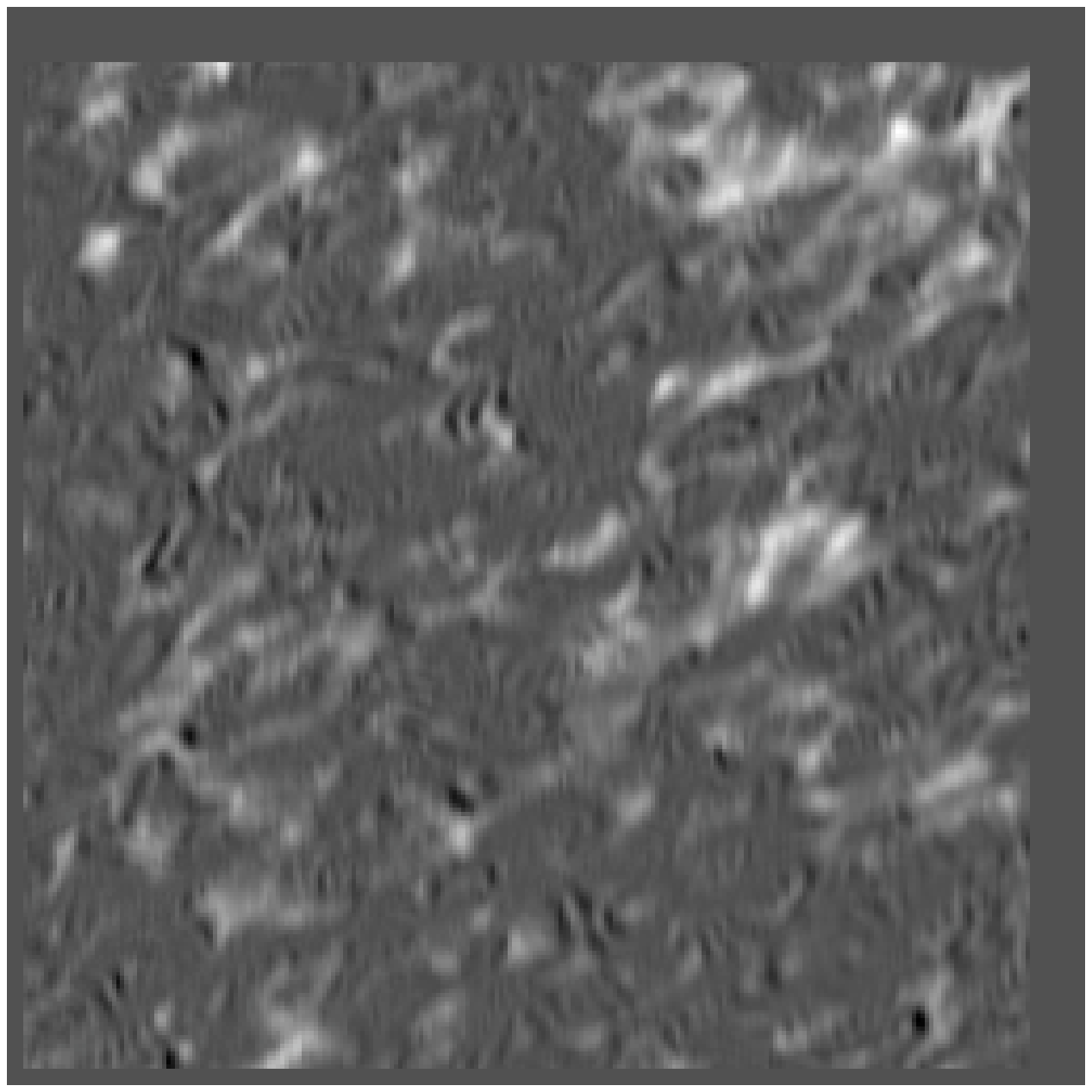}
}
\vskip-5.cm
\caption{\baselineskip=0.5cm{ 
{\bf Projections of the $W_{1}$ DA map and KQ85 mask centred at 
${\rm (long,lat)=\{(0, 45^{\circ}N), (0, 0), (0, 45^{\circ}S)\}}$ respectively first, second and third row, and the resulting convergence patches.}
Left panels: The masked CMB patches.
Central panels: The mask patches, with amplitude 0 on the black regions and amplitude 1 elsewhere. 
Right panels: The $\hat \kappa-\big<\hat \kappa\big>_{\rm pixel}$ patches obtained from 
the corresponding masked CMB patches.
}}
\label{fig:patches}
\end{figure}

We generate the kernel for each DA. In the calculation of the kernel we use the lensed CMB power spectrum measured by WMAP, combined with the window beam function and the detector noise characteristic of the DA. 
For the fiducial unlensed power spectrum we use the power spectrum computed with CAMB. We plot the $m=0$ mode in Fig.~\ref{fig:kernel_da}. All kernels have similar size and similar structures within the  $1\%$ level, differing only in the smaller scale structures (of order $<0.1^{\circ}$)  at sub--percent level.
For the numerical implementation we truncate the kernel at the $1\%$ level, which corresponds to $\theta_{\textrm{kernel}}=1.1^{\circ}$ similarly to the experiments in the previous section.
From each patch $i$ of the WMAP map from the W--band DA $W_j,$ we compute the corresponding convergence map $\hat \kappa_{W_j}^{i}.$ 
For illustration, we show three patches of the $W_1$ DA map obtained from projections centered at the galactic coordinates ${\rm (long,lat)=\{(0, 45^{\circ}N), (0, 0) ,(0, 45^{\circ}S)\}},$ and the corresponding convergence patches (Fig.~\ref{fig:patches}).
The contours of the mask are closely reproduced in the convergence patches, which demonstrates that the reconstruction is local.

\subsection{The convergence power spectrum}

To compute the Gaussian bias, we reconstruct the convergence map from various realizations of the unlensed patch corresponding to the single realization of the observed patch. The average power spectrum of these convergence maps, weighted by the inverse of their variance according to Eqn.~(\ref{eqn:cl_unlensed_w}), is our calculation of the Gaussian contribution of the CMB to the convergence power spectrum of the patch. Thus, for each patch $i$ and for each DA $W_j,$ we have
\ba
\left(C_{\ell}^{\kappa_{\vert\psi}}\right)_{i,W_{j}W_{j^{\prime}}}=
\left(C_{\ell}^{\hat\kappa}\right)_{i,W_{j}W_{j^{\prime}}}-
\left(\big<C_{\ell}^{\hat\kappa_{\vert\psi=0}}\big>_{CMB}\right)_{i,W_{j}W_{j^{\prime}}}.
\label{eqn:cl_wmap_iWjWj}
\ea 
For each patch, we compute the average over the four DA's, which includes four auto--correlations and six cross--correlations in both the $C_{\ell}^{\hat\kappa}$ and the $\big<C_{\ell}^{\hat\kappa_{\vert\psi=0}}\big>_{CMB}$ terms, weighted by the inverse of the variance of each $\left(C_{\ell}^{\kappa_{\vert\psi}}\right)_{i,W_{j}W_{j^{\prime}}}$ which is given by Eqn.~(\ref{eqn:var_cl_lens_w}). 
This operation yields $\left(C_{\ell}^{\kappa_{\vert\psi}}\right)_{i}$ with variance computed by the standard procedure as in Eqn.~(\ref{eqn:var_cl_unlensed_w}).

\begin{figure}[t]
\setlength{\unitlength}{1cm}
\vskip-1.5cm
\centerline{
\hskip-0.5cm
\includegraphics[width=12cm]
{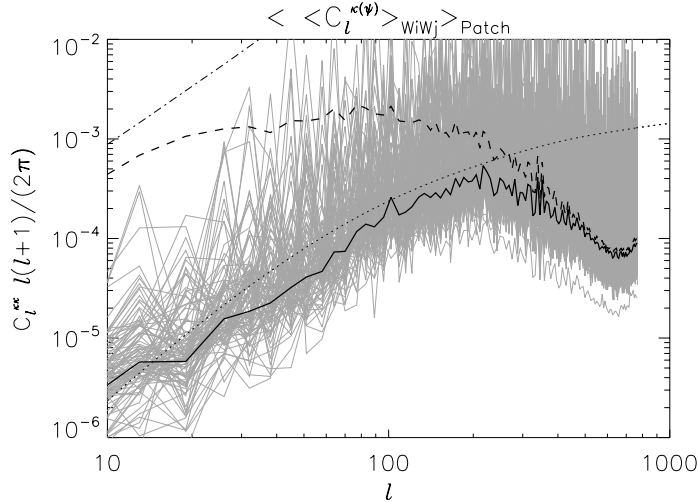}
}
\vskip-7.5cm
\caption{\baselineskip=0.5cm{ 
{\bf  Dispersion of the convergence power spectrum as a function of the lensed CMB realization used to define the bias calibration.} The black dashed line is the raw  convergence power spectrum obtained from the CMB patches after subtracting the convergence power spectrum from unlensed CMB realizations. The black filled line is the convergence power spectrum obtained after calibrating the black dashed line with the realization of Fig.~\ref{fig:cl_kk2_designer_da}, 
whereas the gray lines are the spectra obtained after calibrating with the various lensed CMB realizations.
}}
\label{fig:cl_kk2_cal_cmb}
\end{figure}

We then compute the average over all the patches, weighted by the variance of each $\left(C_{\ell}^{\kappa_{\vert\psi}}\right)_{i}.$ 
We recall that masking a map causes a degradation in the amplitude of the computed convergence power spectrum in comparison to the power spectrum computed from the complete map. This amplitude degradation has been quantified for masks of point sources with varying density and radius \cite{tereno}. (Other methods have been proposed to estimate the bias caused by masking the CMB map \cite{namikawa12,levy13}.) We correct the convergence power spectrum of each patch by the amplitude degradation $A({f_{\rm masked}}_{i})$ inferred for the fractional area of the corresponding mask patch ${f_{\rm masked}}_{i},$ predominantly consisting of contiguous large areas.
We also apply the bias calibration $f_{\rm cal}$ 
derived from the simulations. We recall that the bias calibration allows one to deconvolve the biased convergence power spectrum computed from a single WMAP--like map to obtain the unbiased power spectrum. Hence, the final result is
\ba
C_{\ell}^{\kappa_{\vert\psi}}=
f^{-1}_{\rm cal}~\big<A\left({f_{\rm masked}}_{i}\right)
\left(C_{\ell}^{\kappa_{\vert\psi}}\right)_{i}\big>_{i}.
\ea
The variance of  $C_{\ell}^{\kappa_{\vert\psi}}$ depends on the variances of $f_{\rm cal}$ and $\left(C_{\ell}^{\kappa_{\vert\psi}}\right)_{i}.$
The variance of $f_{\rm cal},$ already shown in Fig.~\ref{fig:cl_kk2_designer_da} (right panel), is computed from the sample dispersion of the different realizations of the lensed CMB map, as illustrated in Fig.~\ref{fig:cl_kk2_cal_cmb}.
 


\begin{figure}[t]
\setlength{\unitlength}{1cm}
\vskip-1.5cm
\centerline{
\hskip-0.5cm
\includegraphics[width=12cm]
{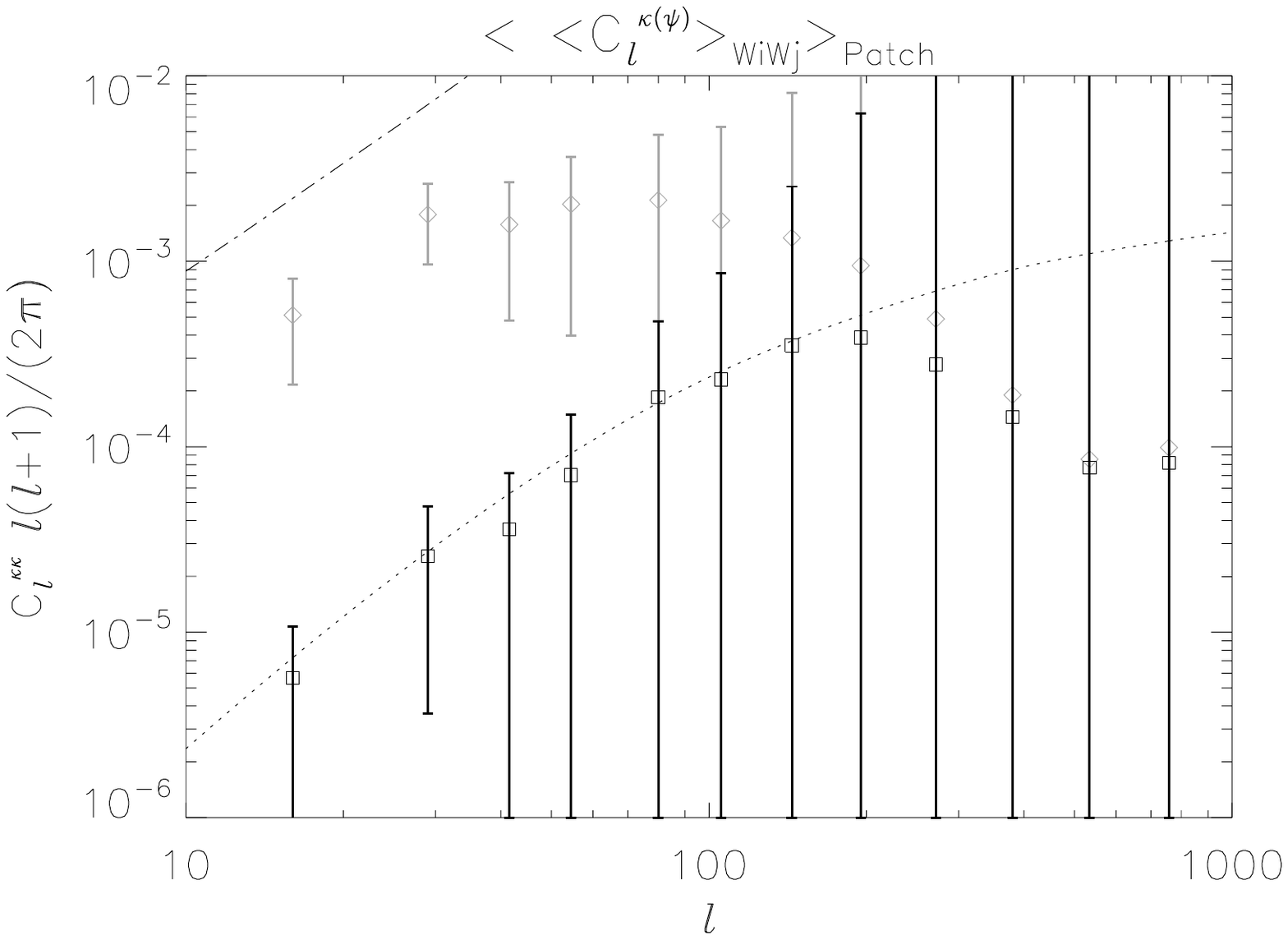}
\hskip-3cm
\includegraphics[width=12cm]
{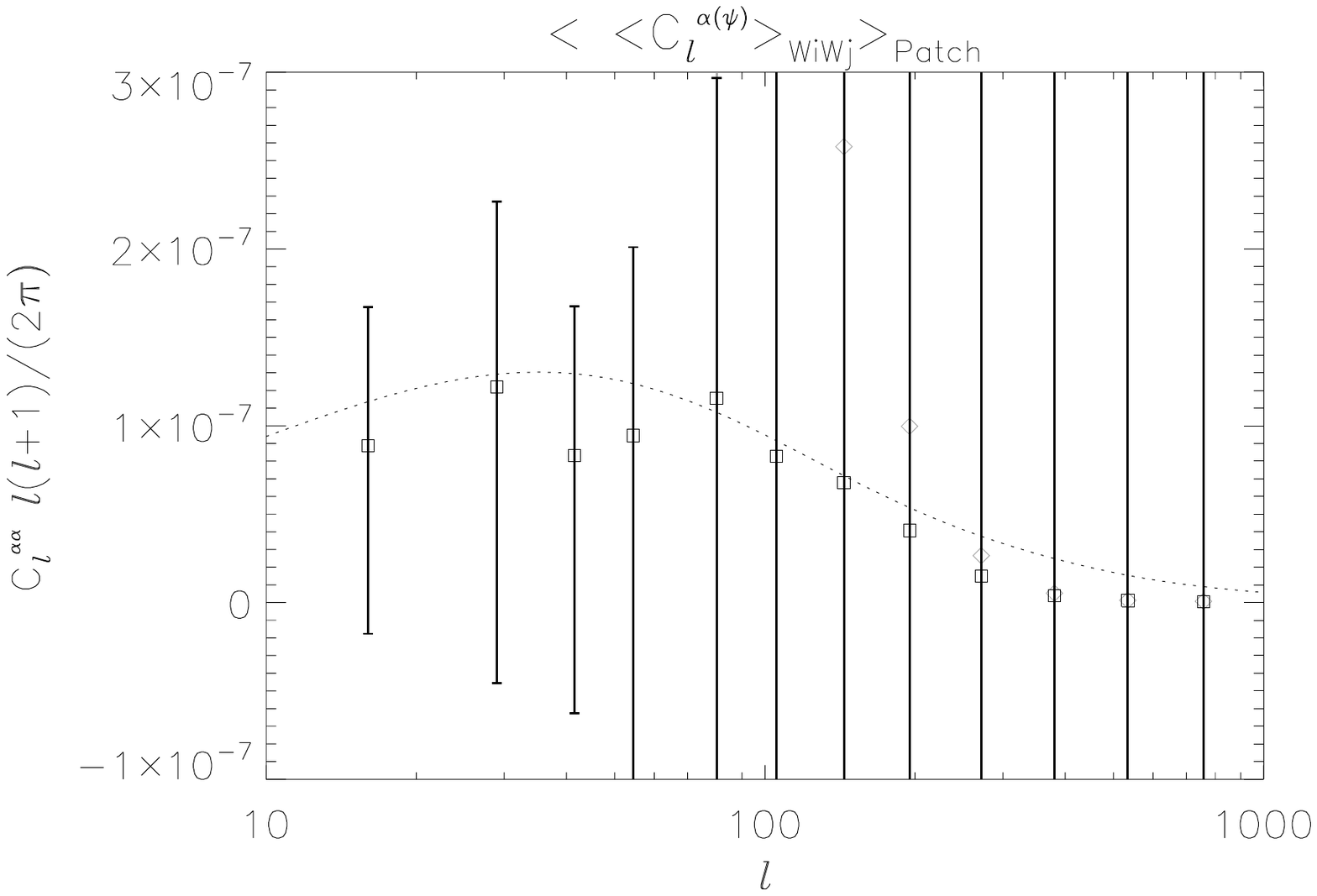}
}
\vskip-7.5cm
\caption{\baselineskip=0.5cm{ 
{\bf 
Convergence power spectrum computed from the CMB maps of the WMAP's W--band DAs.}
The gray diamonds are the weighted average of the difference between the power spectrum measured from the patches and the power spectrum measured from the simulations for each DA and for each patch. The black squares are the same quantity multiplied by the bias calibration. 
The dotted line is the theoretical convergence power spectrum computed with CAMB. The dash--dotted line is the inverse sum of the variance of the estimator ${\rm Var}[\hat \kappa]$ for each DA. Left panel: The convergence power spectrum. Right panel: The deflection power spectrum. 
}}
\label{fig:cl_kk2_da}
\end{figure}

The results, before and after the bias calibration, are presented in Fig.~\ref{fig:cl_kk2_da}, for both the convergence and the deflection field (which is the quantity most commonly used, for easier comparison with the results from other detections). 
After correcting with the bias calibration, 
we recover a lensing signal which agrees with the theoretical prediction. 

In terms of the lensing amplitude parameter, 
the results for the four experiments, separately and combined, in the validity range of the estimator, are presented in Table~\ref{table:A_L_wmap}.
For the combined result we find $A_{L}=0.99\pm 1.67,$ which is similar to the findings in Ref.~\cite{feng11}.  
Since the distribution of calibrations implies an uncertainty in the calibration applied, the error of $A_L$ estimated from the data is larger than that of  $A_L$ estimated from simulations (Sec.~\ref{sec:wmap_ps})
where the calibration was exactly known.
   
\begin{table}[t]
\begin{tabular}{c||cccc|cc}
\hline
~&~$(A_{L}^{\rm obs})_{W_1}$~&~$(A_{L}^{\rm obs})_{W_2}$
~&~$(A_{L}^{\rm obs})_{W_3}$~&~$(A_{L}^{\rm obs})_{W_4}$
~&~$(A_{L}^{\rm obs})_{\left<W_iW_i\right>}$
~&~$(A_{L}^{\rm obs})_{\left<W_iW_j\right>}$
\\ \hline
$\ell \in  [6,198]$
& $1.12\pm5.41$ & $1.07\pm7.09$ & $1.14\pm7.42$ & $1.30\pm6.43$
& $1.16\pm3.22$ & $0.99\pm1.67$
\\ \hline 
\end{tabular}
\vskip.1cm
\caption{\label{table:A_L_wmap} \baselineskip=0.5cm{
{\bf $A_L^{\rm obs}$ from each W--band DA 
and from the combined W--band DAs.} 
}}
\end{table}

\subsection{Consistency tests}

\begin{figure}[t]
\setlength{\unitlength}{1cm}
\vskip-1.5cm
\centerline{
\hskip-0.5cm
\includegraphics[width=12cm]
{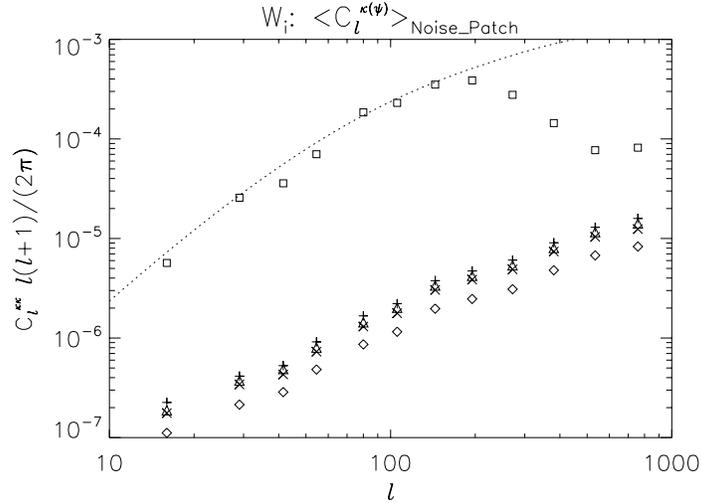}
}
\vskip-7.5cm
\caption{\baselineskip=0.5cm{ 
{\bf 
Convergence power spectrum computed from noise maps of the WMAP's W--band DAs.}
The symbols are the convergence power spectrum averaged over the different patches for each DA. Diamond: $W_{1}.$ Cross: $W_{2}.$ Plus:  $W_{3}.$ Triangle: $W_{4}.$
The squares are the convergence power spectrum measured from the CMB maps (after the bias calibration), which we include for comparison. The dotted line is the theoretical convergence power spectrum computed with CAMB. 
}}
\label{fig:cl_kk2_noise_da}
\end{figure}

First, as a null test, we measure the convergence power spectrum from noise maps. 
In order to do this, we apply the estimator to noise patches synthesised for each DA from the corresponding $\sigma_{0}$ and $N_{\rm obs}.$ These are the same maps that we have added to the simulated unlensed CMB maps (see Subsec.~\ref{subsec:wmap_input_maps}). For each DA, we average the convergence power spectrum over the different patches and present the results in Fig.~\ref{fig:cl_kk2_noise_da}.
 The results are consistent with zero power, showing that the estimator is free of systematics. 
We notice that this result did not require a bias calibration, thus highlighting the lensing origin of the additional non--Gaussian correlations since they are not present in this test on noise maps.

\begin{figure}[t]
\setlength{\unitlength}{1cm}
\centerline{
\hskip-0.5cm
\includegraphics[width=12cm]
{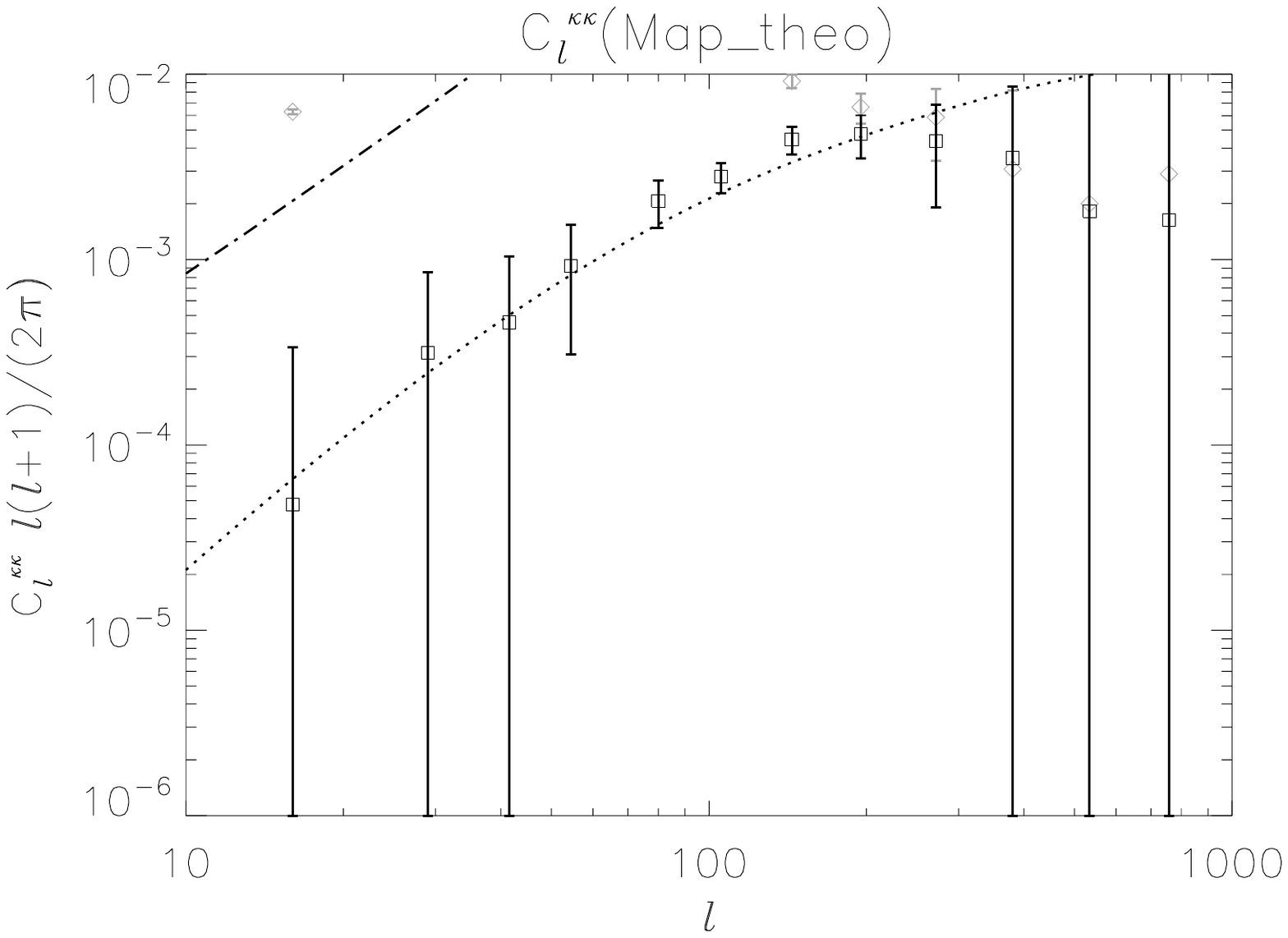}
\hskip-3cm
\includegraphics[width=12cm]
{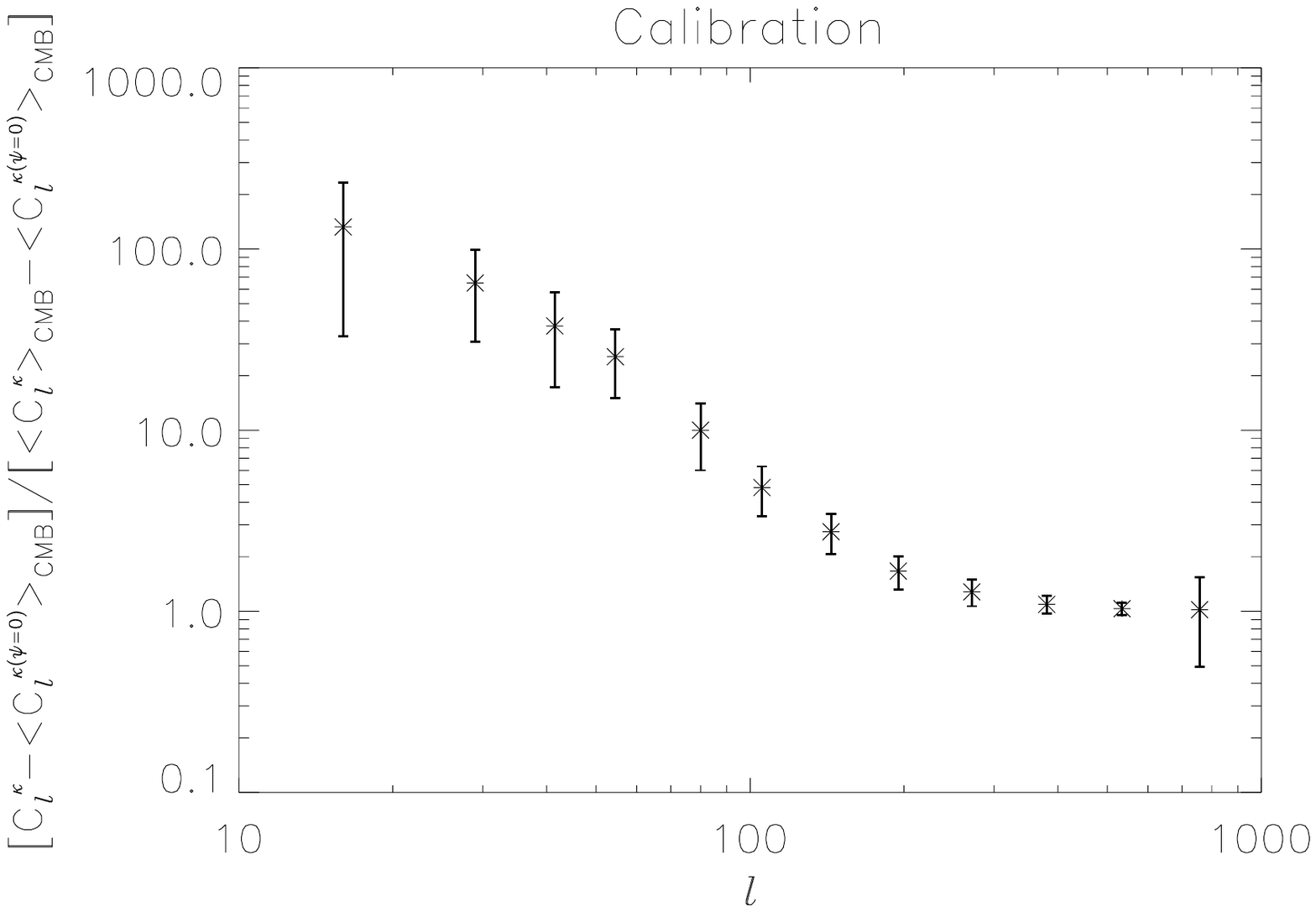}
}
\vskip-7.5cm
\caption{\baselineskip=0.5cm{ 
{\bf 
Convergence power spectrum computed from CMB simulations of the WMAP's W--band DAs with nine times the fiducial lensing power spectrum.}
Left panel: The gray diamonds are the convergence power spectrum computed as the difference between one lensed CMB map and the inverse--variance average over different realizations of unlensed CMB maps. 
The black squares are the convergence power spectrum computed as the difference between the inverse--variance average over different realizations of paired lensed CMB and unlensed CMB maps. For each DA, the maps are synthesised from the theoretical power spectra (with the input lensing power spectrum set to nine times the fiducial one), combined with the beam window function and noise power spectrum of the corresponding DA. The dotted line is the theoretical convergence power spectrum computed with CAMB. The dash--dotted line is the inverse sum of the variance of the estimator ${\rm Var}[\hat \kappa_{\vert\psi}]$ for each DA.
Right panel: The black stars show the bias calibration $f_{\rm cal}$ for $\psi=3~\!\psi_{\rm fid}.$ 
}}
\label{fig:cl_kk2_designer3psi_da}
\end{figure}

We also test the general applicability of the calibration method. In order to do this, we change the input lensing power spectrum (by a constant factor) and compute the corresponding lensed temperature power spectrum. The cosmological parameters and unlensed power spectrum were left unchanged. This case may account for scenarios of modified theories of gravity where lensing properties are different while keeping the standard cosmology.  We thus repeat the procedure in Sec.~\ref{sec:wmap_ps}, namely we produce realizations of the lensed temperature map, add the WMAP noise maps and apply the estimator obtaining the convergence power spectra given by  Eqn.~(\ref{eqn:oldestimator}). Using Eqn.~(\ref{eqn:newestimator}) we obtain an unbiased convergence power spectrum that recovers the input (Fig.~\ref{fig:cl_kk2_designer3psi_da}, left panel). More importantly, the bias calibration factor derived from these simulations is, on all scales, within $1\sigma$ of the calibration factor derived from the fiducial lensing potential (Fig.~\ref{fig:cl_kk2_designer3psi_da}, right panel). This is a good indication that the bias calibration is independent of the amplitude of the lensing potential and can be applied without previous knowledge of the amount of lensing present. 

\section{Conclusions}
\label{sec:wmap_end}

In this manuscript we present a direct detection of the weak lensing from the WMAP 7--year data. We used the real--space variant of the optimal quadratic estimator originally defined in Fourier space.

First we applied the estimator to maps synthesised so as to simulate the WMAP W--band. The conventional estimator of the Gaussian bias failed to recover the input convergence power spectrum, in contrast with its success in Planck simulations. This failure signaled the presence of an additional bias. A variation of this estimator was conceived which successfully accounted for the additional bias and recovered the input convergence power spectrum, with a lensing parameter of $A_{L}=1.01\pm1.46.$ The detection error is dominated by the variance of the estimator which depends on the experimental resolution and detector noise. By comparing the results of the two estimators, we derived an bias calibration which estimates the additional bias as a function of $\ell.$ Moreover, we showed that the bias calibration thus defined is independent of the amplitude of the lensing potential power spectrum.

We proceeded to apply the estimator to maps obtained from HEALPix cartesian projections of the full--sky maps produced by the WMAP W--band. We used the conventional estimator of the Gaussian bias, including the correction of amplitude degradation due to masking, and the bias calibration function. We achieved a good agreement of the detection with the theoretical prediction, as well as with the detection from the simulations, yielding $A_{L}=0.99\pm1.67.$ This detection is also compatible with previous detections. 
However, the large error renders this detection not statistically significant, similarly to the detection from the WMAP data using the harmonic implementation of the optimal estimator \cite{feng11}. The detection from the WMAP data using a kurtosis estimator has higher significance because it aims for the lensing correlations of interest \cite{smidt11}.

Nonetheless, our detection serves as a proof of the robustness of the real--space estimator in dealing with extended masks and overwhelming experimental noise. Moreover, by comparing the results of a WMAP simulation with those of a Planck simulation, we bring attention to the dependence of biases on the experiment resolution.
We expect that a full--sky experiment with higher resolution and lower detector noise, such as Planck \cite{planck11}, will allows us to make a more significant measurement of the convergence power spectrum.\\


\noindent{\bf Acknowledgments} CSC is funded by Funda\c{c}\~ao para a Ci\^encia e a Tecnologia (FCT), Grant no. SFRH/BPD/65993/2009.  IT is funded by FCT, Grant no. SFRH/BPD/65122/2009 and acknowledges support from the European Programme FP7-PEOPLE-2010-RG-268312. CSC and IT also acknowledge support from the project CERN/FP/123618/2011. The authors acknowledge the use of the HEALPix package for the handling of the full--sky maps.
The authors thank C. Feng, M.G.R. Santos, D.N. Spergel and E. Vagenas for useful discussions. The authors also thank the referees for comments that helped to improve the manuscript.

\end{document}